\documentstyle[12pt]{article}
\newcommand{\Li}{{\mathrm{Li}_2}}
\newcommand{\dd}{{\mathrm{d}}}

\title{Table of integrals. Asymptotical
expressions for non--collinear kinematics.}

\author{A.B. Arbuzov, A.V. Belitsky, E.A. Kuraev, and B.G. Shaikhatdenov }

\date{}

\begin{document}

\maketitle

\begin{center}
{\it Joint Institute for Nuclear Research, Dubna, 141980, Russia }
\end{center}

\begin{abstract}
We present a set of Feynman integrals appearing in calculations
of different QED processes to the one--loop accuracy.
We consider scalar, vector, and tensor integrals with two, three,
four and five denominators. The cases of equal and different fermion
masses are considered. 
Results obtained are valid in the region where all kinematical invariants
are large compared to the masses squared. Mass corrections for some
scalar integrals in the case of different fermion masses are also given.
\end{abstract}

\section{Introduction}

This paper is an electronic version of our preprint~\cite{preprint}
from year 1998. Since that time the Table of integrals was used by
our group in several calculations of 1-loop QED radiative corrections
to various processes at high energues. It might be useful also for some
more applications, estimates and cross checks.

We give below a list of 4--dimensional integrals in momentum space
which were used in calculation of 1--loop radiative corrections to the
inelastic processes of kind $e^+e^-\to e^+e^-\gamma$ and
$\mu^-e^-\to \mu^-e^-\gamma$ at high energies.
For definiteness we consider the case when all the scalar products
of external 4--momenta are large compared to the mass of external real
particles squared $p_ip_j \gg p_i^2=m_i^2$. We omit systematically
terms of order $m_i^2/p_ip_j$ compared to ones of order unity and
restrict ourselves to consideration of scattering type Feynman diagrams
only. Remaining ones can be obtained by application of crossing
and analytical continuation transformations. We introduce the ultraviolet,
infrared cut--offs and do not use the dimensional regularization.
In paper of one of authors ~\cite{kur80} 1--loop integrals, related
to the $2 \to 2$ type Feynman diagrams, were considered. Here we give
a set of integrals for description of inelastic $2 \to 3$ type processes.
Our paper is organized as follows. In the first part we consider the
4--momentum integrals, appearing in one--loop vertex and five--point
diagrams. The cases when fermions have different and equal masses are considered.
They may be helpful in constructing the five--point diagrams with one or two
external off--shell mass particles . Scalar, vector and tensor integrals are
considered up to the case of four denominators. Tensor integrals are
calculated up to a third rank.

\section{Five--point Feynman diagrams}
\begin{center}
{ \large \bf The case of equal fermion masses.}
\end{center}
Typical process: $\qquad e^+e^-\to e^+e^-\gamma$

\subsection{Notations}

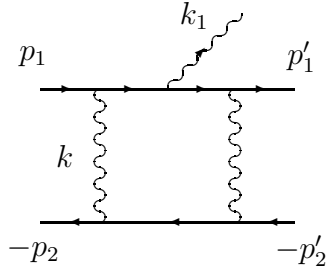
\begin{figure}[h]
\unitlength=2.10pt
\special{em:linewidth 0.4pt}
\linethickness{0.4pt}
\begin{picture}(131.66,51.67)
\put(119.57,47.68){\oval(2.67,2.67)[lt]}
\put(119.57,50.34){\oval(2.67,2.67)[rb]}
\put(116.90,45.01){\oval(2.67,2.67)[lt]}
\put(116.90,47.68){\oval(2.67,2.67)[rb]}
\put(114.23,42.34){\oval(2.67,2.67)[lt]}
\put(114.23,45.01){\oval(2.67,2.67)[rb]}
\put(111.57,39.68){\oval(2.67,2.67)[lt]}
\put(111.57,42.34){\oval(2.67,2.67)[rb]}
\put(108.90,37.01){\oval(2.67,2.67)[lt]}
\put(108.90,39.68){\oval(2.67,2.67)[rb]}
\put(113.22,43.23){\vector(1,1){1.11}}
\put(95.33,13.67){\oval(2.00,2.00)[r]}
\put(95.33,15.67){\oval(2.00,2.00)[l]}
\put(95.33,17.67){\oval(2.00,2.00)[r]}
\put(95.33,19.67){\oval(2.00,2.00)[l]}
\put(95.33,21.67){\oval(2.00,2.00)[r]}
\put(95.33,23.67){\oval(2.00,2.00)[l]}
\put(95.33,25.67){\oval(2.00,2.00)[r]}
\put(95.33,27.67){\oval(2.00,2.00)[l]}
\put(95.33,29.67){\oval(2.00,2.00)[r]}
\put(95.33,31.67){\oval(2.00,2.00)[l]}
\put(95.33,33.67){\oval(2.00,2.00)[r]}
\put(95.33,35.67){\oval(2.00,2.00)[l]}
\put(119.66,13.67){\oval(2.00,2.00)[r]}
\put(119.66,15.67){\oval(2.00,2.00)[l]}
\put(119.66,17.67){\oval(2.00,2.00)[r]}
\put(119.66,19.67){\oval(2.00,2.00)[l]}
\put(119.66,21.67){\oval(2.00,2.00)[r]}
\put(119.66,23.67){\oval(2.00,2.00)[l]}
\put(119.66,25.67){\oval(2.00,2.00)[r]}
\put(119.66,27.67){\oval(2.00,2.00)[l]}
\put(119.66,29.67){\oval(2.00,2.00)[r]}
\put(119.66,31.67){\oval(2.00,2.00)[l]}
\put(119.66,33.67){\oval(2.00,2.00)[r]}
\put(119.66,35.67){\oval(2.00,2.00)[l]}
\put(84.66,36.67){\line(1,0){45.67}}
\put(84.66,12.67){\line(1,0){45.67}}
\put(89.33,36.67){\vector(1,0){1.00}}
\put(100.67,36.67){\vector(1,0){1.00}}
\put(113.67,36.67){\vector(1,0){1.00}}
\put(124.33,36.67){\vector(1,0){1.00}}
\put(91.00,12.67){\vector(-1,0){1.33}}
\put(109.00,12.67){\vector(-1,0){1.33}}
\put(126.67,12.67){\vector(-1,0){1.33}}
\put(83.33,42.67){\makebox(0,0)[cc]{$p_1$}}
\put(131.66,42.67){\makebox(0,0)[cc]{$p_1'$}}
\put(131.66,7.67){\makebox(0,0)[cc]{$-p_2'$}}
\put(83.33,7.67){\makebox(0,0)[cc]{$-p_2$}}
\put(89.00,24.67){\makebox(0,0)[cc]{$k$}}
\put(112.33,50.00){\makebox(0,0)[cc]{$k_1$}}
\end{picture}
\caption{\bf Five--point Feynman diagram}
\end{figure}

\begin{equation}
I^{1,\mu,\mu\nu}_{ijklm}
\equiv\int\frac{\dd^4k}{i\pi^2}\frac{1,k^{\mu},k^{\mu}k^{\nu}}{(i)(j)(k)(l)(m)}
\end{equation}

\begin{eqnarray}
&(1)=(p_1-k)^2-m^2,\ (2)=(p_1-k_1-k)^2-m^2,\ (3)=(p_2+k)^2-m^2,  \\ \nonumber
&(4)=(q-k)^2-\lambda^2,\quad (5)=k^2-\lambda^2.
\end{eqnarray}

Invariants
\begin{eqnarray}
&\chi_1=2p_1k_1,\ \chi_1'=2p_1'k_1,\ \chi_2=2p_2k_1=s-s_1-\chi_1,\nonumber\\
&\chi_2'=2p_2'k_1=s-s_1-\chi_1',\ s=(p_1+p_2)^2,\ s_1=(p_1'+p_2')^2,\nonumber\\
&t=q^2,\ q=p_2'-p_2,\ t_1=q'^2=t+\chi_1-\chi_1',\ q'=p_1'-p_1,\nonumber\\
&\chi_1+\chi_2=s-s_1,\ p_1^2=p_2^2=p_1'^2=p_2'^2=m^2,\ k_1^2=0.
\end{eqnarray}

\begin{eqnarray}
&L_t=\ln\left(\frac{-t}{m^2}\right),\hspace{2mm}
L_s=\ln\left(\frac{s}{m^2}\right),\hspace{2mm}
L_{\Lambda}=\ln\left(\frac{\Lambda^2}{m^2}\right),\hspace{2mm}
L_{\lambda}=\ln\left(\frac{\lambda^2}{m^2}\right),\hspace{2mm} \\ \nonumber
&\Li(z)=-\int_{0}^{z}\frac{\dd x}{x}\ln(1-x),
\end{eqnarray}
\begin{equation}
{\cal P}^2=m^2-x\bar x s-i0,\hspace{0.3cm}
{\cal P}_1^2=m^2-x\bar x s_1-i0,
\hspace{0.3cm}\mbox{where}\hspace{0.3cm} \bar x=1-x
\end{equation}
Throughout the paper $\Lambda$ is an UV cut--off parameter and
$\lambda$ is a fictitious {\it photon mass}.

\subsection{Two-propagator integrals}
\subsubsection{Scalar integrals}

\begin{eqnarray}
I_{12}&=&-1+L_{\Lambda},\quad
I_{13}=-1-\int_{0}^{1}\dd x\ln\left(\frac{{\cal P}^2}
{\Lambda^2}\right)=1+L_{\Lambda}-L_{s}+i\pi, \nonumber\\
I_{14}&=&-\int_{0}^{1}\dd x\ln\left(\frac{xm^2-\bar x \chi_1'}{\Lambda^2}\right)
=1+L_{\Lambda}-L_{\chi_1'}+i\pi,\nonumber\\
I_{15}&=&I_{24}=I_{34}=I_{35}=1+L_{\Lambda},\
I_{23}=-1-\int_{0}^{1}\dd x\ln\left(\frac{{\cal P}_1^2}{\Lambda^2}\right)
=1+L_{\Lambda}-L_{s_1}+i\pi,\nonumber\\
I_{25}&=&-\int_{0}^{1}\dd x\ln\left(\frac{xm^2+\bar x \chi_1}
{\Lambda^2}\right)=1+L_{\Lambda}-L_{\chi_1},\quad I_{45}=1+L_{\Lambda}-L_t.
\end{eqnarray}

\subsubsection{Vector integrals}

\begin{eqnarray}
I^{\mu}_{12}&=&\left(p_1-\frac{k_1}{2}\right)^{\mu}
\left(L_{\Lambda}-\frac{3}{2}\right),\
I^{\mu}_{13}=\left(p_1-p_2\right)^{\mu}
\left(\frac{1}{4}+\frac{1}{2}L_{\Lambda}-
\frac{1}{2}L_{s}+\frac{i\pi}{2}\right),\nonumber\\
I^{\mu}_{14}&=&\left(p_1+q\right)^{\mu}
\left(\frac{1}{4}+\frac{1}{2}L_{\Lambda}-
\frac{1}{2}L_{\chi_1'}+\frac{i\pi}{2}\right),\
I^{\mu}_{15}=\frac{p_1^{\mu}}{2}\left(L_{\Lambda}-\frac{1}{2}\right),\nonumber\\
I^{\mu}_{24}&=&\left(p_1-k_1\right)^{\mu}
\left(\frac{1}{2}+L_{\Lambda}\right)-\frac{p_1'^{\mu}}{2}
\left(\frac{3}{2}+L_{\Lambda}\right),\nonumber\\
I^{\mu}_{34}&=&\frac{p_2'^{\mu}}{2}\left(\frac{3}{2}+L_{\Lambda}\right)-
p_2^{\mu}\left(\frac{1}{2}+L_{\Lambda}\right),\nonumber\\
I^{\mu}_{35}&=&\frac{p_2^{\mu}}{2}
\left(\frac{1}{2}-L_{\Lambda}\right),\
I^{\mu}_{23}=\left(p_1-k_1-p_2\right)^{\mu}
\left(\frac{1}{4}+\frac{1}{2}L_{\Lambda}-
\frac{1}{2}L_{s_1}+\frac{i\pi}{2}\right),\nonumber\\
I^{\mu}_{25}&=&\left(p_1-k_1\right)^{\mu}
\left(\frac{1}{4}+\frac{1}{2}L_{\Lambda}-
\frac{1}{2}L_{\chi_1}\right),\
I^{\mu}_{45}=q^{\mu}\left(\frac{1}{4}+\frac{1}{2}L_{\Lambda}-
\frac{1}{2}L_t\right).
\end{eqnarray}

\subsection{Three-propagator integrals}
\subsubsection{Scalar integrals}

\begin{eqnarray}
I_{123}&=&\frac{1}{s-s_1}\int_{0}^{1}\frac{\dd x}{x}
\ln\left(\frac{{\cal P}^2}{{\cal P}_1^2}\right)
=\frac{1}{s-s_1}\left[\frac{1}{2}L_{s}^2
-\frac{1}{2}L_{s_1}^2+i\pi\left( L_{s_1}-L_{s} \right)\right],\nonumber\\
I_{345}&=&-\int_{0}^{1}\frac{\dd x}{{\bar x}^2m^2+xt}
\ln\left(\frac{m^2{\bar x}^2}{-tx}\right)
=\frac{1}{t}\left[\frac{1}{2}L_{t}^2 +4\zeta (2)\right],\nonumber\\
I_{124}&=&\int_{0}^{1}\frac{\dd x\dd y}{x\bar y \chi_1'-ym^2}
=\frac{1}{\chi_1'}\left[\frac{1}{2}L_{\chi_1'}^2
-\zeta (2)-i\pi L_{\chi_1'}\right],\nonumber\\
I_{125}&=&-\int_{0}^{1}\frac{\dd x\dd y}{x\bar y \chi_1 +ym^2}
=\frac{1}{\chi_1}\left[-\frac{1}{2}L_{\chi_1}^2
-2\zeta (2)\right],\nonumber\\
I_{134}&=&-\int_{0}^{1}\frac{\dd x\dd y}{y {\cal P}^2-x\bar y \chi_1'}
=\frac{1}{s-\chi_1'}\biggl[\frac{3}{2}L_{s}^2+\frac{1}{2}L_{\chi_1'}^2
-2L_{s}L_{\chi_1'}+2\Li\left(1-\frac{\chi_1'}{s}\right) \nonumber\\
&+& i\pi\left( L_{\chi_1'}-L_{s} \right)\biggr],\nonumber\\
I_{235}&=&-\int_{0}^{1}\frac{\dd x\dd y}{y {\cal P}_1^2+x\bar y \chi_1}
=\frac{1}{s_1+\chi_1}\biggl[\frac{3}{2}L_{s_1}^2+\frac{1}{2}L_{\chi_1}^2
-2L_{s_1}L_{\chi_1}-9\zeta (2) \nonumber\\
&+&2\Li\left(1+\frac{\chi_1}{s_1}\right)
+i\pi\left( 2L_{\chi_1}-3L_{s_1} \right)\biggr],\nonumber\\
I_{135}&=&-\frac{1}{2}\int_{0}^{1}\frac{\dd x}{{\cal P}^2}
\ln\left(\frac{{\cal P}^2}{\lambda^2}\right)
=\frac{1}{s}\left[\frac{1}{2}L_{s}^2-L_{s}L_{\lambda}-4\zeta (2)
+i\pi\left( L_{\lambda}-L_{s} \right)\right],\nonumber\\
I_{234}&=&-\frac{1}{2}\int_{0}^{1}\frac{\dd x}{{\cal P}_1^2}
\ln\left(\frac{{\cal P}_1^2}{\lambda^2}\right)
=\frac{1}{s_1}\left[\frac{1}{2}L_{s_1}^2-L_{s_1}L_{\lambda}-4\zeta (2)
+i\pi\left( L_{\lambda}-L_{s_1} \right)\right],\nonumber\\
I_{245}&=&\int_{0}^{1}\frac{\dd x}{x\chi_1-\bar xt-x^2m^2}
\ln\left(\frac{x^2m^2}{x\chi_1-\bar xt}\right)
=\frac{1}{\chi_1+t}\biggl[\frac{1}{2}L_{t}^2-\frac{1}{2}L_{\chi_1}^2 \nonumber\\
&+&2\Li\left(1-\frac{\chi_1}{-t}\right)\biggr],\nonumber\\
I_{145}&=&-\int_{0}^{1}\frac{\dd x\dd y}{y{\bar x}^2m^2
-\bar yxt- yx\bar x\chi_1'}
=\frac{1}{\chi_1'-t}\biggl[\frac{1}{2}L_{\chi_1'}^2-\frac{1}{2}L_{t}^2
-3\zeta (2) \nonumber\\
&-&2\Li\left(1+\frac{\chi_1'}{-t}\right)-i\pi L_{\chi_1'}\biggr].\nonumber\\
\end{eqnarray}

\subsubsection{Vector integrals}
The parameterization reads
\begin{equation}
I^{\mu}_{ijk}=a_{ijk}p_1^{\mu}+b_{ijk}p_2^{\mu}
+c_{ijk}k_1^{\mu}+d_{ijk}p_1'^{\mu}.
\end{equation}

\begin{eqnarray}
a_{245}&=&-c_{245}=\frac{1}{t+\chi_1}\left[
\frac{1}{2}L_{t}^2-\frac{1}{2}L_{\chi_1}^2+L_{\chi_1}-L_t
+2\Li\left(1-\frac{\chi_1}{-t}\right)\right],\ b_{245}=0,\nonumber\\
d_{245}&=&\frac{1}{t+\chi_1}\left[ -L_t
-\frac{\chi_1}{t+\chi_1}
\left[ \frac{1}{2}L_{t}^2-\frac{1}{2}L_{\chi_1}^2+2L_{\chi_1}-2L_t
+2\Li\left(1-\frac{\chi_1}{-t}\right)\right]
\right].\\
&&\nonumber\\
a_{145}&=&\frac{1}{\chi_1'-t}\biggl[ 2L_{\chi_1'}-L_t-2i\pi
+\frac{t}{\chi_1'-t}\biggl[\frac{1}{2}L_{t}^2-\frac{1}{2}L_{\chi_1'}^2
\nonumber\\
&+&2L_{\chi_1'}-2L_t +3\zeta (2)+2\Li\left(1+\frac{\chi_1'}{-t}\right)
+i\pi \left(L_{\chi_1'}-2\right) \biggr]\biggr],\nonumber\\
b_{145}&=&0,\ c_{145}=d_{145}=\frac{1}{\chi_1'-t}
\left[ L_t - L_{\chi_1'}+i\pi\right].\\
&&\nonumber\\
a_{345}&=&-c_{345}=-d_{345}=\frac{1}{t}L_t,\ b_{345}
=\frac{1}{t}\left[ -\frac{1}{2}L_t^2+2L_t-4\zeta (2)\right].\\
&&\nonumber\\
a_{125}&=&\frac{1}{\chi_1}\left[-\frac{1}{2}L_{\chi_1}^2+L_{\chi_1}
-2\zeta (2)\right],\ b_{125}=d_{125}=0,\ c_{125}=\frac{1}{\chi_1}
\left[ L_{\chi_1}-2 \right].\\
&&\nonumber\\
a_{235}&=&-c_{235}=\frac{1}{s_1+\chi_1}\left[ L_{s_1}-L_{\chi_1}
-i\pi \right],\ d_{235}=0,\nonumber\\
b_{235}&=&\frac{1}{s_1+\chi_1}
\biggl[ -L_{s_1}+i\pi\nonumber\\
&+&\frac{\chi_1}{s_1+\chi_1}\biggl[ 2L_{s_1}-2L_{\chi_1}
-\frac{3}{2}L_{s_1}^2-\frac{1}{2}L_{\chi_1}^2
+2L_{s_1}L_{\chi_1}+9\zeta (2)
-2\Li\left(1+\frac{\chi_1}{s_1}\right)\nonumber\\
&+&i\pi\left(-2- 2L_{\chi_1}+3L_{s_1}\right) \biggr]\biggr]. \\
&&\nonumber\\
a_{135}&=&-b_{135}=\frac{1}{s}\left[ L_s-i\pi\right],\ c_{135}=d_{135}=0.\\
&&\nonumber\\
a_{234}&=&-c_{234}=-b_{234}-d_{234}=\frac{1}{s_1}\biggl[
\frac{1}{2}L_{s_1}^2-L_{s_1}L_{\lambda}-L_{s_1}-4\zeta (2) \nonumber\\
&+&i\pi\left(1+ L_{\lambda}-L_{s_1} \right)\biggr],
b_{234}=a_{234}-I_{234}=\frac{1}{s_1}\left[-L_{s_1} +i\pi\right],\nonumber\\
d_{234}&=&\frac{1}{s_1}\left[
-\frac{1}{2}L_{s_1}^2+L_{s_1}L_{\lambda}+2L_{s_1}+4\zeta (2)
+i\pi\left(-2 + L_{s_1} -  L_{\lambda} \right)\right].\\
&&\nonumber\\
a_{134}&=&\frac{1}{s-\chi_1'}\left[
L_{s}-2L_{\chi_1'}+i\pi
-\frac{2s}{s-\chi_1'}\left[ L_{s}-L_{\chi_1'} \right]
+sI_{134}\right],\nonumber\\
b_{134}&=&a_{134}-I_{134}=\frac{1}{s-\chi_1'}\left[-L_{s}+i\pi
-\frac{2\chi_1'}{s-\chi_1'}\left[ L_{s}-L_{\chi_1'} \right]
+\chi_1'I_{134}\right],\nonumber\\
c_{134}&=&d_{134}=\frac{1}{s-\chi_1'}\left[L_{\chi_1'}-i\pi
+\frac{2s}{s-\chi_1'}\left[ L_{s}-L_{\chi_1'} \right]
-sI_{134}\right].\\
&&\nonumber\\
a_{124}&=&I_{124}=\frac{1}{\chi_1'}\left[\frac{1}{2}L_{\chi_1'}^2
-\zeta (2)-i\pi L_{\chi_1'}\right],\ d_{124}=\frac{1}{\chi_1'}
\left[-L_{\chi_1'} +i\pi \right], \nonumber\\
c_{124}&=&\frac{1}{\chi_1'}\left[
-\frac{1}{2}L_{\chi_1'}^2 + L_{\chi_1'} + \zeta (2)-2
+i\pi \left( L_{\chi_1'}-1 \right)\right],\ b_{124}=0. \\
&&\nonumber\\
a_{123}&=&b_{123}-I_{123}=\frac{1}{s-s_1}\left[
\frac{1}{2}L_{s}^2 -\frac{1}{2}L_{s_1}^2
-L_s+L_{s_1} +i\pi\left( L_{s_1}-L_{s} \right)
\right],\nonumber\\
b_{123}&=&-\frac{1}{s-s_1}\left[ L_s-L_{s_1} \right],\ d_{123}=0,\
c_{123}=\frac{1}{s-s_1}\biggl[L_{s_1}-2-i\pi \nonumber\\
&+&\frac{s}{s-s_1}\left[-\frac{1}{2}L_{s}^2 +\frac{1}{2}L_{s_1}^2
+2L_s-2L_{s_1} +i\pi\left( L_{s}-L_{s_1} \right)\right]\biggr].\nonumber\\
\end{eqnarray}

\subsection{Four-propagator integrals}
\subsubsection{Scalar integrals}
\begin{eqnarray}
I_{1245}&=&\int_{0}^{1}\frac{\dd x \dd y}{\left[ xy\chi_1-\bar yt\right]
\left[ ym^2-\bar x\bar y \chi_1'\right]}=\frac{1}{\chi_1\chi_1'}\biggl[
-L_{\chi_1}^2-L_{\chi_1'}^2-L_t^2-2L_{\chi_1}L_{\chi_1'} \nonumber \\
&+&2L_{\chi_1}L_t+2L_{\chi_1'}L_t+4\zeta (2)
+i\pi\left( 2L_{\chi_1}+2L_{\chi_1'}-2L_t \right)\biggr],\nonumber\\
I_{2345}&=&\frac{1}{t}\int_{0}^{1}\frac{\dd x}{{\cal P}_1^2}\left[
-\frac{1}{2}\ln\left(\frac{{\cal P}_1^2}{\lambda^2}\right)
+\ln\left( \frac{x\chi_1}{-t} \right)\right]
=\frac{1}{s_1t}\biggl[
L_{s_1}^2-L_{s_1}L_{\lambda}-2L_{s_1}L_{\chi_1} \nonumber\\
&+&2L_{s_1}L_t-5\zeta (2)+i\pi\left( 2L_{\chi_1}-2L_t-2L_{s_1}
+L_{\lambda} \right) \biggr],\nonumber\\
I_{1345}&=&\frac{1}{t}\int_{0}^{1}\frac{\dd x}{{\cal P}^2}\left[
-\frac{1}{2}\ln\left(\frac{{\cal P}^2}{\lambda^2}\right)
+\ln\left( \frac{-x\chi_1'}{-t} \right)\right]\nonumber\\
&=&\frac{1}{st}\left[
L_{s}^2-L_{s}L_{\lambda}-2L_{s}L_{\chi_1'}+2L_{s}L_t+7\zeta (2)
+i\pi\left( 2L_{\chi_1'}-2L_t+L_{\lambda} \right)
\right],\nonumber\\
I_{1235}&=&\frac{1}{\chi_1}\int_{0}^{1}\frac{\dd x}{{\cal P}^2}\left[
\frac{1}{2}\ln\left(\frac{{\cal P}^2}{\lambda^2}\right)
-\ln\left( \frac{{\cal P}_1^2}{x\chi_1} \right)\right]
=\frac{1}{s\chi_1}\biggl[
L_{s_1}^2-2L_{s}L_{\chi_1}+L_{s}L_{\lambda}\nonumber\\
&-&5\zeta (2)+2\Li\left( 1-\frac{s_1}{s}\right)
+i\pi\left( -2L_{s_1}+2L_{\chi_1}-L_{\lambda} \right)
\biggr],\nonumber\\
I_{1234}&=&\frac{1}{\chi_1'}\int_{0}^{1}\frac{\dd x}{{\cal P}_1^2}\left[
-\frac{1}{2}\ln\left(\frac{{\cal P}_1^2}{\lambda^2}\right)
+\ln\left( \frac{{\cal P}^2}{-x\chi_1'} \right)\right]
=\frac{1}{s_1\chi_1'}\biggl[
-L_{s}^2+2L_{s_1}L_{\chi_1'}    \nonumber\\
&-&L_{s_1}L_{\lambda}-7\zeta (2)-2\Li\left( 1-\frac{s}{s_1}\right)
+i\pi\left( 2L_{s}-2L_{s_1}-2L_{\chi_1'}+L_{\lambda} \right)
\biggr].
\end{eqnarray}

Useful integrals
\begin{eqnarray}
\int_{0}^{1}\frac{\dd x}{{\cal P}^2}&=&\frac{2}{s}\left[-L_s+i\pi\right],\
\int_{0}^{1}\frac{\dd x}{{\cal P}^2}\ln x
=\frac{1}{s}\left[\frac{1}{2}L_s^2-\zeta (2)-i\pi L_s\right],\nonumber\\
\int_{0}^{1}\frac{\dd x}{{\cal P}^2}\ln \left( \frac{{\cal P}^2}{m^2}\right)
&=&\frac{1}{s}\left[-L_s^2+8\zeta (2)+2i\pi L_s\right],\nonumber\\
\int_{0}^{1}\frac{\dd x}{{\cal P}^2}\ln \left( \frac{{\cal P}_1^2}{m^2}\right)
&=&\frac{1}{s}\left[-L_{s_1}^2+8\zeta (2)-2\Li\left(1-\frac{s_1}{s} \right)
+2i\pi L_{s_1}\right].
\end{eqnarray}

\subsubsection{Vector integrals}

Parameterization
\begin{equation}
I^{\mu}_{ijkl}=a_{ijkl}p_1^{\mu}+b_{ijkl}p_2^{\mu}
+c_{ijkl}k_1^{\mu}+d_{ijkl}p_1'^{\mu}
\end{equation}

\begin{equation}
a_{1245}=\frac{\Delta^{(1)}}{\Delta},\hspace{0.3cm}
b_{1245}=0,\hspace{0.3cm}
c_{1245}=\frac{\Delta^{(2)}}{\Delta},\hspace{0.3cm}
d_{1245}=\frac{\Delta^{(3)}}{\Delta},\hspace{0.3cm}
\end{equation}
\begin{eqnarray}
\Delta&=&-2t_1\chi_1\chi_1',\quad \Delta^{(1)}=\chi_1'
\biggl[ -\chi_1'I_{124} + \chi_1I_{125} + (\chi_1 + t)I_{245} \nonumber\\
&+&(\chi_1' - t -2\chi_1 )I_{145} + \chi_1 (\chi_1'
- 2(t +\chi_1) )I_{1245} \biggr],\nonumber\\
\Delta^{(2)}&=&t_1\left[ -\chi_1'I_{124} - \chi_1I_{125}
+ (\chi_1 + t)I_{245}
+ (\chi_1' - t )I_{145} + \chi_1 \chi_1' I_{1245} \right],\nonumber\\
\Delta^{(3)}&=&\chi_1
\left[ \chi_1'I_{124} - \chi_1I_{125} + (\chi_1 + t -2\chi_1' )I_{245}
+ (\chi_1' - t )I_{145} + \chi_1 \chi_1' I_{1245} \right].
\end{eqnarray}

\begin{equation}
a_{1235}=\frac{\Delta^{(1)}}{\Delta},\hspace{0.3cm}
b_{1235}=\frac{\Delta^{(2)}}{\Delta},\hspace{0.3cm}
c_{1235}=\frac{\Delta^{(3)}}{\Delta},\hspace{0.3cm}
d_{1235}=0,\hspace{0.3cm}
\end{equation}
\begin{eqnarray}
\Delta&=&-2s\chi_1\chi_2,\nonumber\\
\Delta^{(1)}&=&\chi_2
\left[ (\chi_1+\chi_2)I_{123} - \chi_1I_{125} -sI_{135}
+ (s- \chi_2 )I_{235}-s \chi_1 I_{1235} \right], \nonumber\\
\Delta^{(2)}&=&\chi_1
\left[ -(\chi_1+\chi_2)I_{123} + \chi_1I_{125} -sI_{135}
+ (s+ \chi_2 )I_{235}-s \chi_1 I_{1235} \right],\nonumber\\
\Delta^{(3)}&=&s \left[ (\chi_1-\chi_2)I_{123} - \chi_1I_{125}+sI_{135}
- (s- \chi_2 )I_{235} +s \chi_1 I_{1235}\right].
\end{eqnarray}

\begin{equation}
a_{1345}=\frac{\Delta^{(1)}}{\Delta},\hspace{0.3cm}
b_{1345}=\frac{\Delta^{(2)}}{\Delta},\hspace{0.3cm}
c_{1345}=d_{1345}=\frac{\Delta^{(3)}}{\Delta},\hspace{0.3cm}
\end{equation}
\begin{eqnarray}
u&=&-s-t+\chi_1',\ \Delta=2stu,\ \Delta^{(1)}=
-(-su+t\chi_1')I_{134} + s(s+t)I_{135} \nonumber\\
&+&t(s+t)I_{345}-(-tu+s\chi_1')I_{145}-st(s+t) I_{1345},\ \Delta^{(2)}=
-(-\chi_1'u+st)I_{134} \nonumber\\
&-&s(t-\chi_1')I_{135}+t(2s+t-\chi_1')I_{345}
-(t-\chi_1')^2I_{145}+st(t-\chi_1') I_{1345},\nonumber\\
\Delta^{(3)}&=&s\left[ ( s+2t-\chi_1')I_{134} - sI_{135} -tI_{345}
-(t-\chi_1')I_{145}+st I_{1345} \right].
\end{eqnarray}

\begin{equation}
a_{2345}=-c_{2345}=\frac{\Delta^{(1)}}{\Delta},\hspace{0.3cm}
b_{2345}=\frac{\Delta^{(2)}}{\Delta},\hspace{0.3cm}
d_{2345}=\frac{\Delta^{(3)}}{\Delta},\hspace{0.3cm}
\end{equation}
\begin{eqnarray}
u_1&=&-s_1-t-\chi_1,\ \Delta=2s_1tu_1,\nonumber\\
\Delta^{(1)}&=&-u_1\left[ -(t+\chi_1)I_{245} - s_1I_{234}
+(s_1+\chi_1)I_{235}+tI_{345} -s_1t I_{2345}\right],\nonumber\\
\Delta^{(2)}&=&-(t+\chi_1)^2I_{245} - s_1(t+\chi_1)I_{234}
+(-\chi_1u_1-s_1t)I_{235} \nonumber\\
&+&t(2s_1+t+\chi_1)I_{345}+s_1t(t+\chi_1) I_{2345},\nonumber\\
\Delta^{(3)}&=&(-\chi_1u_1
-s_1t)I_{245}+ s_1(s_1+2t+\chi_1)I_{234}-(s_1+\chi_1)^2I_{235} \nonumber\\
&-&t(s_1+\chi_1)I_{345}+s_1t(s_1+\chi_1) I_{2345}.
\end{eqnarray}

\begin{eqnarray}
a_{1234}&=&I_{1234}+\frac{\Delta^{(2)}}{\Delta},\
b_{1234}=\frac{\Delta^{(2)}}{\Delta},\
c_{1234}=-I_{1234}-\frac{\Delta^{(2)}}{\Delta
}+\frac{\Delta^{(3)}}{\Delta},\nonumber\\
d_{1234}&=&-I_{1234}+\frac{\Delta^{(1)}}{\Delta}
-\frac{\Delta^{(2)}}{\Delta},
\end{eqnarray}
\begin{eqnarray}
\Delta&=&2s_1\chi_1'\chi_2',\hspace{0.3cm}\chi_2'=s-s_1-\chi_1',\nonumber\\
\Delta^{(1)}&=&\chi_2'
\left[ -(s-s_1)I_{123} + (s-\chi_1')I_{134} + \chi_1'I_{124}
- s_1I_{234} + s_1\chi_1'I_{1234} \right],\nonumber\\
\Delta^{(2)}&=&\chi_1'
\left[ (s-s_1)I_{123} + (2s_1-s+\chi_1')I_{134} - \chi_1'I_{124}
- s_1I_{234} + s_1\chi_1'I_{1234} \right],\nonumber\\
\Delta^{(3)}&=&s_1
\left[ (\chi_2'-\chi_1')I_{123} - (s-\chi_1')I_{134} + \chi_1'I_{124}
+ s_1I_{234} - s_1\chi_1'I_{1234} \right].
\end{eqnarray}

\subsubsection{Tensor}

Parameterization

\begin{eqnarray}
I^{\mu\nu}_{ijkl}&=&g^T_{ijkl}g^{\mu\nu}+a^T_{ijkl}p_1^{\mu}p_1^{\nu}
+b^T_{ijkl}p_2^{\mu}p_2^{\nu}+c^T_{ijkl}k_1^{\mu}k_1^{\nu}
+d^T_{ijkl}p_1'^{\mu}p_1'^{\nu}\nonumber\\
&+&\alpha^T_{ijkl}\{ p_1^{\mu}p_2^{\nu} \}
+\beta^T_{ijkl}\{ p_1^{\mu}k_1^{\nu} \}
+\gamma^T_{ijkl}\{ p_1^{\mu}p_1'^{\nu} \}\nonumber\\
&+&\rho^T_{ijkl}\{ p_2^{\mu}p_1'^{\nu} \}
+\sigma^T_{ijkl}\{ p_2^{\mu}k_1^{\nu} \}
+\tau^T_{ijkl}\{ p_1'^{\mu}k_1^{\nu} \},
\end{eqnarray}
where $\{\cdots\}$ means symmetrization with respect to Lorentz indices:
$\{v_{\mu}u_{\nu}\}=v_{\mu}u_{\nu}+v_{\nu}u_{\mu}$.

\begin{eqnarray}
g^T_{1245}
&=&\frac{1}{2}\left[ 2I_{124}-a_{124}-\chi_1 c_{1245} +(t+\chi_1)d_{1245}
\right],\nonumber\\
a^T_{1245}&=&\frac{1}{t_1\chi_1}
\biggl[ \chi_1'(-I_{124}+a_{124}-c_{145})+t_1a_{145} -(t+\chi_1)a_{245}
\nonumber\\
&+&t_1\chi_1a_{1245} -\chi_1'(t+\chi_1)d_{1245} \biggr], \nonumber\\
c^T_{1245}&=&\frac{1}{\chi_1\chi_1'}
\left[ t_1(-I_{124}+a_{124})+\chi_1c_{125}+(t_1-\chi_1)c_{145}
-\chi_1\chi_1'c_{1245} \right],\nonumber\\
d^T_{1245}&=&\frac{1}{t_1\chi_1'}
\left[ \chi_1(-I_{124}+a_{124}-a_{245})+(t_1-\chi_1)c_{145}
-t_1d_{245} -\chi_1\chi_1'd_{1245} \right],\nonumber\\
\beta^T_{1245}&=&\frac{1}{\chi_1}
\left[ -I_{124} +a_{124}+c_{145}+\chi_1c_{1245}\right],\
b^T_{1245}=\alpha^T_{1245}= \rho^T_{1245}= \sigma^T_{1245}=0, \nonumber\\
\gamma^T_{1245}&=&\frac{1}{t_1}
\left[ I_{124}-a_{124}+a_{245}+c_{145}+(t+\chi_1)d_{1245}\right],\nonumber\\
\tau^T_{1245}&=&\frac{1}{\chi_1'}
\left[ -I_{124}+a_{245}+\chi_1c_{1245}-(t+\chi_1)d_{1245}\right].
\end{eqnarray}

\begin{eqnarray}
g^T_{1235}&=&\frac{1}{2}
\left[ 2I_{123}-a_{123}+b_{123} -\chi_1c_{1235} \right],\nonumber\\
a^T_{1235}&=&\frac{1}{s\chi_1}
\left[ \chi_2I_{123}-(\chi_1+\chi_2)a_{123}
+\chi_1a_{125}-\chi_1\chi_2c_{1235} \right],\nonumber\\
b^T_{1235}&=&\frac{1}{s\chi_2}
\left[ \chi_1(I_{123}-a_{235})+(\chi_1+\chi_2)b_{123}
-\chi_2b_{235}-\chi_1^2c_{1235} \right],\nonumber\\
c^T_{1235}&=&\frac{1}{\chi_1\chi_2}
\left[ s(I_{123}+b_{123})-(s-\chi_2)a_{235}
+\chi_2c_{123}-s\chi_1c_{1235} \right],\ d^T_{1235}=0,\nonumber\\
\alpha^T_{1235}&=&\frac{1}{s}
\left[ -I_{123} +a_{123}-a_{235}-b_{123} \right],\
\beta^T_{1235}=\frac{1}{\chi_1}
\left[ -I_{123} +a_{123}+\chi_1c_{1235} \right],\nonumber\\
\sigma^T_{1235}&=&\frac{1}{\chi_2}
\left[ -I_{123} +a_{235}-b_{123}+\chi_1c_{1235} \right],\
\gamma^T_{1235} = \rho^T_{1235} = \tau^T_{1235} =0.
\end{eqnarray}

\begin{eqnarray}
g^T_{1345}&=&\frac{1}{2}\left[ I_{134}+tc_{1345} \right],\
b^T_{1345}=\frac{1}{s}
\left[ b_{134}-b_{345}-(\chi_1'-t)\rho^T_{1345} \right],\nonumber\\
a^T_{1345}&=&\frac{1}{st(\chi_1'-s-t)}
[ (s+t)^2 I_{134}+t(\chi_1'-s-t)a_{145}-(s(s+t)+t\chi_1')a_{134}\nonumber\\
&&\hspace{4cm}+\chi_1'(s+t)(c_{145}-c_{134})+t(s+t)^2c_{1345} ],\nonumber\\
c^T_{1345}&=&d^T_{1345}=\tau^T_{1345}
=\frac{1}{t(\chi_1'-s-t)}
\left[ (\chi_1'-t)(c_{145}-c_{134})-s(b_{134}-tc_{1345}) \right],\nonumber\\
\alpha^T_{1345}&=&\frac{1}{st(\chi_1'-s-t)}
[ -t(\chi_1'-s-t)a_{345}+\chi_1'(\chi_1'-t)(c_{145}-c_{134})\nonumber\\
&-&s\chi_1'(a_{134}-I_{134})+st\chi_1'c_{1345} ],\nonumber\\
\beta^T_{1345}&=&\gamma^T_{1345}=\frac{1}{t(\chi_1'-s-t)}
\left[ (s+t)(b_{134}-tc_{1345})-\chi_1'(c_{145}-c_{134})\right],\nonumber\\
\rho^T_{1345}&=&\sigma^T_{1345}=\frac{1}{st(\chi_1'-s-t)}
[ (\chi_1'(\chi_1'-t)-st)c_{134}-(\chi_1'-t)^2c_{145} \nonumber\\
&+&t(\chi_1'-s-t)a_{345}+s(\chi_1'-t)b_{134}-st(\chi_1'-t)c_{1345}].
\end{eqnarray}

\begin{eqnarray}
g^T_{2345}&=&\frac{1}{2}
\left[ I_{234}+\chi_1a_{2345}+(t+\chi_1)d_{2345} \right],\
\alpha^T_{2345}=-\sigma^T_{2345}=\frac{1}{s_1t}[-\chi_1a_{235}\nonumber\\
&-&ta_{345}],\ a^T_{2345}=c^T_{2345}=-\beta^T_{2345}=\frac{1}{s_1t}
\left[ -ta_{345}-(s_1+\chi_1)a_{235}+s_1ta_{2345} \right],\nonumber\\
b^T_{2345}&=&\frac{1}{s_1t(\chi_1+s_1+t)}
[ s_1t(b_{235}-b_{345})-\chi_1(t+\chi_1)a_{235} \nonumber\\
&-&t(t+\chi_1)a_{345}-s_1t(t+\chi_1)b_{2345}],\nonumber\\
d^T_{2345}&=&\frac{1}{(\chi_1+s_1+t)}
\biggl[ d_{245}-d_{234}-\frac{(\chi_1+s_1)}{s_1t(\chi_1+s_1+t)}
\biggl[ s_1t(a_{245}-a_{234}) \nonumber\\
&+&t(\chi_1+s_1)a_{345}+(\chi_1+s_1)^2a_{235}
-s_1t(\chi_1+s_1)a_{2345}\biggr] \biggr],\nonumber\\
\gamma^T_{2345}&=&-\tau^T_{2345}=\frac{1}{s_1t(\chi_1+s_1+t)}
[ s_1t(a_{245}-a_{234})+t(\chi_1+s_1)a_{345} \nonumber\\
&+&(\chi_1+s_1)^2 a_{235}-s_1t(\chi_1+s_1)a_{2345} ],\nonumber\\
\rho^T_{2345}&=&\frac{1}{s_1t(\chi_1+s_1+t)}
[ -s_1ta_{234} +\chi_1(\chi_1+s_1)a_{235}+t(\chi_1+s_1)a_{345} \nonumber\\
&-&s_1t\chi_1a_{2345}-s_1t(\chi_1+t)d_{2345}].
\end{eqnarray}

\begin{eqnarray}
g^T_{1234}&=&\frac{1}{2}
\left[ I_{123}-\chi_1'\frac{\Delta^{(3)}}{\Delta} \right],\
a^T_{1234}=
2\frac{\Delta^{(2)}}{\Delta} +I_{1234}+\tilde b_{1234},\
b^T_{1234}=\tilde b_{1234},\nonumber\\
c^T_{1234}
&=&2\frac{\Delta^{(2)}}{\Delta}- 2\frac{\Delta^{(3)}}{\Delta} +I_{1234}
+\tilde b_{1234}+\tilde c_{1234}-2\tilde\gamma_{1234},\nonumber\\
d^T_{1234}
&=&2\frac{\Delta^{(2)}}{\Delta}- 2\frac{\Delta^{(1)}}{\Delta} +I_{1234}
+\tilde a_{1234}+\tilde b_{1234}-2\tilde\alpha_{1234},\nonumber\\
\alpha^T_{1234}
&=&\frac{\Delta^{(2)}}{\Delta}+\tilde b_{1234},\
\beta^T_{1234}
=\frac{\Delta^{(3)}}{\Delta}- 2\frac{\Delta^{(2)}}{\Delta} -I_{1234}
-\tilde b_{1234}+\tilde\gamma_{1234},\nonumber\\
\gamma^T_{1234}
&=&\frac{\Delta^{(1)}}{\Delta}- 2\frac{\Delta^{(2)}}{\Delta} -I_{1234}
-\tilde b_{1234}+\tilde\alpha_{1234},\nonumber\\
\rho^T_{1234}&=&-\frac{\Delta^{(2)}}{\Delta}
-\tilde b_{1234}+\tilde\alpha_{1234},\
\sigma^T_{1234}=-\frac{\Delta^{(2)}}{\Delta}
-\tilde b_{1234}+\tilde\gamma_{1234},\nonumber\\
\tau^T_{1234}
&=&2\frac{\Delta^{(2)}}{\Delta}- \frac{\Delta^{(1)}}{\Delta}
-\frac{\Delta^{(3)}}{\Delta} +I_{1234}+\tilde b_{1234}-\tilde\alpha_{1234}
+\tilde\beta_{1234}-\tilde\gamma_{1234},
\end{eqnarray}
where
\begin{eqnarray}
\tilde a_{1234}&=&\frac{1}{s_1\chi_1'}\left[\chi_1'(I_{124}-I_{123}+d_{124})
-(\chi_1'+\chi_2')b_{123}-\chi_1'\chi_2'\frac{\Delta^{(3)}}{\Delta}
\right],\nonumber\\
\tilde b_{1234}&=&\frac{1}{s_1\chi_2'}\left[
-\chi_1'(I_{134}-I_{123}+c_{134})+(\chi_1'+\chi_2')(b_{123}-b_{134})
-\chi_1'^2\frac{\Delta^{(3)}}{\Delta}\right],\nonumber\\
\tilde c_{1234}&=&\frac{1}{\chi_1'\chi_2'}\biggl[
(s_1+\chi_2')(I_{123}-I_{134}+b_{123}-b_{134}-c_{134})+\chi_2'c_{123}\nonumber\\
&-&s_1\chi_1'\frac{\Delta^{(3)}}{\Delta}\biggr],\nonumber\\
\tilde\alpha_{1234}&=&\frac{1}{s_1}\left[ -b_{134}-c_{134}-I_{134} \right],\
\tilde\beta_{1234}=\frac{1}{\chi_1'}
\left[  b_{123}+\chi_1'\frac{\Delta^{(3)}}{\Delta}\right],\nonumber\\
\tilde\gamma_{1234}&=&\frac{1}{\chi_2'}
\left[ -b_{123}+b_{134}+c_{134}+I_{134}
-I_{123}+\chi_1'\frac{\Delta^{(3)}}{\Delta}\right].
\end{eqnarray}

\subsubsection{Pentagon}
Following ref.~\cite{ver84} we express the pentagon diagram in terms of
box graphs.
\begin{equation}
I_{12345}=-\frac{1}{\Delta}\left[
\Delta^{(1)}I_{2345}+\Delta^{(2)}I_{1345}+\Delta^{(3)}I_{1245}
+\Delta^{(4)}I_{1235}+\Delta^{(5)}I_{1234}\right],
\end{equation}
where
\begin{eqnarray}
&&\Delta=2ss_1t\chi_1\chi_1',\ \Delta^{(1)}
=s_1t\left[-(s-s_1)t-s \chi_1 -s_1\chi_1'-\chi_1\chi_1'\right],\nonumber\\
&&\Delta^{(2)}
=st\left[(s-s_1)t+s \chi_1 +s_1\chi_1'-\chi_1\chi_1'\right],\nonumber\\
&&\Delta^{(3)}=\chi_1\chi_1'
\left[-(s+s_1)t-s \chi_1 +s_1\chi_1'+\chi_1\chi_1'\right],\nonumber\\
&&\Delta^{(4)}
=s\chi_1\left[(s-s_1)t+s \chi_1 -s_1\chi_1'-\chi_1\chi_1'\right],\nonumber\\
&&\Delta^{(5)}
=s_1\chi_1'\left[(s-s_1)t-s \chi_1 +s_1\chi_1'+\chi_1\chi_1'\right].
\end{eqnarray} \\

\begin{center}
{ \bf The case of different fermion masses.}
\end{center}
Typical process: $\qquad \mu^-e^-\to \mu^-e^-\gamma$

\subsection{Notations (see fig.1)}

\begin{equation}
I^{1,\mu,\mu\nu}_{ijklm}
\equiv\int\frac{\dd^4k}{i\pi^2}\frac{1,k^{\mu},k^{\mu}k^{\nu}}{(i)(j)(k)(l)(m)}
\end{equation}

\begin{eqnarray}
&(1)=(p_1-k)^2-m^2,\quad (2)=(p_1-k_1-k)^2-m^2,\quad (3)=(p_2+k)^2-\mu^2,\nonumber\\
&(4)=(q-k)^2-\lambda^2,\qquad (5)=k^2-\lambda^2.
\end{eqnarray}

Invariants
\begin{eqnarray}
&\chi_1=2p_1k_1,\ \chi_1'=2p_1'k_1,\ \chi_2=2p_2k_1=s-s_1-\chi_1,\nonumber\\
&\chi_2'=2p_2'k_1=s-s_1-\chi_1',\ s=(p_1+p_2)^2,\ s_1=(p_1'+p_2')^2,\nonumber\\
&t=q^2,\hspace{2mm}q=p_2'-p_2,\ t_1=q'^2=t+\chi_1-\chi_1',\ q'=p_1'-p_1,\nonumber\\
&\chi_1+\chi_2=s-s_1,\ p_1^2=p_1'^2=m^2,\ p_2^2=p_2'^2=\mu^2,\ k_1^2=0.
\end{eqnarray}

\begin{eqnarray}
&&L_{\Lambda_m}=\ln\left(\frac{\Lambda^2}{m^2}\right),\hspace{2mm}
L_{\Lambda_\mu}=\ln\left(\frac{\Lambda^2}{\mu^2}\right),\hspace{2mm}
L_{\lambda_m}=\ln\left(\frac{\lambda^2}{m^2}\right),\hspace{2mm}
L_{\lambda_\mu}=\ln\left(\frac{\lambda^2}{\mu^2}\right),\hspace{2mm} \nonumber\\
&&L_{t_m}=\ln\left(\frac{-t}{m^2}\right),\hspace{2mm}
L_{t_\mu}=\ln\left(\frac{-t}{\mu^2}\right),\hspace{2mm}
L_{s_m}=\ln\left(\frac{s}{m^2}\right),\hspace{2mm}
L_{s_\mu}=\ln\left(\frac{s}{\mu^2}\right),\hspace{2mm} \nonumber\\
&&\mbox{Li}_2(z)=-\int_{0}^{z}\frac{\dd x}{x}\ln(1-x).
\end{eqnarray}
\begin{equation}
{\cal P}^2=m^2x+\mu^2(1-x)-x\bar x s-i0,\hspace{0.3cm}
{\cal P}_1^2=m^2x+\mu^2(1-x)-x\bar x s_1-i0,
\hspace{0.3cm}\mbox{where}\hspace{0.3cm} \bar x=1-x
\end{equation}

\subsection{Two-propagator integrals}
\subsubsection{Scalar}

\begin{eqnarray}
I_{12}&=&-1+L_{\Lambda_m},\ I_{13}=1+L_{\Lambda_\mu}-L_{s_\mu}+i\pi,\nonumber\\
I_{14}&=&1+L_{\Lambda_m}-L_{\chi_{1m}'}+i\pi,\
I_{15}=I_{24}=1+L_{\Lambda_m},\nonumber\\
I_{34}&=&I_{35}=1+L_{\Lambda_\mu},\
I_{23}=1+L_{\Lambda_\mu}-L_{s_{1\mu}}+i\pi,\nonumber\\
I_{25}&=&1+L_{\Lambda_m}-L_{\chi_{1m}},\ I_{45}=1+L_{\Lambda_m}-L_{t_m}.
\end{eqnarray}

\subsubsection{Vector}

\begin{eqnarray}
I^{\mu}_{12}&=&\left(p_1-\frac{k_1}{2}\right)^{\mu}
\left(L_{\Lambda_m}-\frac{3}{2}\right),\
I^{\mu}_{13}=\left(p_1-p_2\right)^{\mu}
\left(\frac{1}{4}+\frac{1}{2}L_{\Lambda_m}-
\frac{1}{2}L_{s_m}+\frac{i\pi}{2}\right),\nonumber\\
I^{\mu}_{14}&=&\left(p_1+q\right)^{\mu}
\left(\frac{1}{4}+\frac{1}{2}L_{\Lambda_m}-
\frac{1}{2}L_{\chi_{1m}'}+\frac{i\pi}{2}\right),\
I^{\mu}_{15}=\frac{p_1^{\mu}}{2}
\left(L_{\Lambda_m}-\frac{1}{2}\right),\nonumber\\
I^{\mu}_{24}&=&\left(p_1-k_1\right)^{\mu}
\left(\frac{1}{2}+L_{\Lambda_m}\right)-
\frac{p_1'^{\mu}}{2}\left(\frac{3}{2}+L_{\Lambda_m}\right),\nonumber\\
I^{\mu}_{34}&=&\frac{p_2'^{\mu}}{2}\left(\frac{3}{2}+L_{\Lambda_\mu}\right)-
p_2^{\mu}\left(\frac{1}{2}+L_{\Lambda_\mu}\right),\
I^{\mu}_{35}=\frac{p_2^{\mu}}{2}\left(\frac{1}{2}-
L_{\Lambda_\mu}\right),\nonumber\\
I^{\mu}_{23}&=&\left(p_1-k_1-p_2\right)^{\mu}
\left(\frac{1}{4}+\frac{1}{2}L_{\Lambda_m}-
\frac{1}{2}L_{s_{1m}}+\frac{i\pi}{2}\right),\nonumber\\
I^{\mu}_{25}&=&\left(p_1-k_1\right)^{\mu}
\left(\frac{1}{4}+\frac{1}{2}L_{\Lambda_m}-
\frac{1}{2}L_{\chi_{1m}}\right),\
I^{\mu}_{45}=q^{\mu}\left(\frac{1}{4}+\frac{1}{2}L_{\Lambda_m}-
\frac{1}{2}L_{t_m}\right).
\end{eqnarray}

\subsection{Three-propagator integrals}
\subsubsection{Scalar}

\begin{eqnarray}
I_{123}&=&\frac{1}{s-s_1}\left[\frac{1}{2}L_{s_m}^2
-\frac{1}{2}L_{s_{1m}}^2+i\pi\left( L_{s_{1m}}-L_{s_m} \right)\right],\
I_{345}=\frac{1}{t}\left[\frac{1}{2}L_{t_\mu}^2 +4\zeta (2)\right],\nonumber\\
I_{124}&=&\frac{1}{\chi_1'}\left[\frac{1}{2}L_{\chi_{1m}'}^2
-\zeta (2)-i\pi L_{\chi_{1m}'}\right],\
I_{125}=\frac{1}{\chi_1}\left[-\frac{1}{2}L_{\chi_{1m}}^2
-2\zeta (2)\right],\nonumber\\
I_{134}&=&\frac{1}{s-\chi_1'}\left[\frac{3}{2}L_{s_\mu}^2
+\frac{1}{2}L_{\chi_{1\mu}'}^2-2L_{s_\mu}L_{\chi_{1\mu}'}
+2\mbox{Li}_2\left(1-\frac{\chi_1'}{s}\right)
+i\pi\left( L_{\chi_{1\mu}'}-L_{s_\mu} \right)\right],\nonumber\\
I_{235}&=&\frac{1}{s_1+\chi_1}\biggl[\frac{3}{2}L_{s_{1\mu}}^2
+\frac{1}{2}L_{\chi_{1\mu}}^2-2L_{s_{1\mu}}L_{\chi_{1\mu}}-9\zeta (2) \nonumber\\
&+&2\mbox{Li}_2\left(1+\frac{\chi_1}{s_1}\right)
+i\pi\left( 2L_{\chi_{1\mu}}-3L_{s_{1\mu}}\right)\biggr] \ , \nonumber\\
I_{135}&=&\frac{1}{s}\biggl[\frac{1}{2}L_{s_m}^2-L_{\lambda_m}L_{s_m}
+i\pi\left(L_{\lambda_m}-L_{s_m}\right)-4\zeta (2)
+2\ln^2\frac{\mu}{m}+2\ln\frac{\mu}{m}L_{\lambda_\mu} \biggr] \ , \nonumber\\
I_{234}&=&\frac{1}{s_1}\biggl[\frac{1}{2}L_{s_{1m}}^2-L_{\lambda_m}L_{s_{1m}}
+i\pi\left(L_{\lambda_m}-L_{s_{1m}}\right)-4\zeta (2)
+2\ln^2\frac{\mu}{m} \nonumber\\
&+&2\ln\frac{\mu}{m}L_{\lambda_\mu} \biggr] \ , \nonumber\\
I_{245}&=&\frac{1}{\chi_1+t}\left[\frac{1}{2}L_{t_m}^2
-\frac{1}{2}L_{\chi_{1m}}^2+2\mbox{Li}_2\left(1-\frac{\chi_1}{-t}\right)
\right],\nonumber\\
I_{145}&=&\frac{1}{\chi_1'-t}\left[\frac{1}{2}L_{\chi_{1m}'}^2
-\frac{1}{2}L_{t_m}^2-3\zeta (2)-2\mbox{Li}_2\left(1+\frac{\chi_1'}{-t}\right)
-i\pi L_{\chi_{1m}'}\right].
\end{eqnarray}

\subsubsection{Vector}
Parameterization
\begin{equation}
I^{\mu}_{ijk}=a_{ijk}p_1^{\mu}+b_{ijk}p_2^{\mu}
+c_{ijk}k_1^{\mu}+d_{ijk}p_1'^{\mu}
\end{equation}

\begin{eqnarray}
a_{245}&=&-c_{245}=\frac{1}{t+\chi_1}\left[
\frac{1}{2}L_{t_m}^2-\frac{1}{2}L_{\chi_{1m}}^2+L_{\chi_{1m}}-L_{t_m}
+2\mbox{Li}_2\left(1-\frac{\chi_1}{-t}\right)\right],\nonumber\\
b_{245}&=&0,\
d_{245}=\frac{1}{t+\chi_1}\biggl[ -L_{t_m}-\frac{\chi_1}{t+\chi_1}
\biggl[ \frac{1}{2}L_{t_m}^2-\frac{1}{2}L_{\chi_{1m}}^2
+2L_{\chi_{1m}}-2L_{t_m} \nonumber\\
&+& 2\mbox{Li}_2\left(1-\frac{\chi_1}{-t}\right)\biggr]\biggr].\\
&&\nonumber\\
a_{145}&=&\frac{1}{\chi_1'-t}\biggl[ 2L_{\chi_{1m}'}-L_{t_m}-2i\pi
+\frac{t}{\chi_1'-t}\biggl[
\frac{1}{2}L_{t_m}^2-\frac{1}{2}L_{\chi_{1m}'}^2+2L_{\chi_{1m}'}\nonumber\\
&-&2L_{t_m} +3\zeta (2)+2\mbox{Li}_2\left(1+\frac{\chi_1'}{-t}\right)
+i\pi \left(L_{\chi_{1m}'}-2\right)\biggr]\biggr],\nonumber\\
b_{145}&=&0,\ c_{145}=d_{145}=\frac{1}{\chi_1'-t}
\left[ L_{t_m} - L_{\chi_{1m}'}+i\pi\right].\\
&&\nonumber\\
a_{345}&=&-c_{345}=-d_{345}=\frac{1}{t}L_{t_\mu},\ b_{345}=\frac{1}{t}\left[
-\frac{1}{2}L_{t_\mu}^2+2L_{t_\mu}-4\zeta (2)\right].\\
&&\nonumber\\
a_{125}&=&\frac{1}{\chi_1}\left[-\frac{1}{2}L_{\chi_{1m}}^2+L_{\chi_{1m}}
-2\zeta (2)\right],\ b_{125}=d_{125}=0,\nonumber\\
c_{125}&=&\frac{1}{\chi_1}\left[ L_{\chi_{1m}}-2 \right].\\
&&\nonumber\\
a_{235}&=&-c_{235}=\frac{1}{s_1+\chi_1}\left[ L_{s_{1\mu}}
-L_{\chi_{1\mu}}-i\pi \right], \ d_{235}=0,\nonumber\\
b_{235}&=&\frac{1}{s_1+\chi_1}\biggl[-L_{s_{1\mu}}+i\pi
+\frac{\chi_1}{s_1+\chi_1}\biggl[2L_{s_{1\mu}}-2L_{\chi_{1\mu}}
-\frac{3}{2}L_{s_{1\mu}}^2-\frac{1}{2}L_{\chi_{1\mu}}^2 \nonumber\\
&+&2L_{s_{1\mu}}L_{\chi_{1\mu}}+9\zeta (2)
-2\mbox{Li}_2\left(1+\frac{\chi_1}{s_1}\right)
+i\pi\left(-2- 2L_{\chi_{1\mu}}+3L_{s_{1\mu}}\right)\biggr]\biggr].\\
&&\nonumber\\
a_{135}&=&-b_{135}=\frac{1}{s}\left[ L_{s_m}-i\pi\right],\
c_{135}=d_{135}=0.\\
&&\nonumber\\
a_{234}&=&-c_{234}=-b_{234}-d_{234}
=\frac{1}{s_1}\left[-L_{s_{1\mu}}+i\pi+s_1I_{234}\right],\nonumber\\
b_{234}&=&a_{234}-I_{234}=\frac{1}{s_1}\left[-L_{s_{1\mu}} +i\pi\right],\nonumber\\
d_{234}&=&\frac{1}{s_1}\left[L_{s_{1m}}+L_{s_{1\mu}}-2i\pi-s_1I_{234}\right].\\
&&\nonumber\\
a_{134}&=&\frac{1}{s-\chi_1'}\left[L_{s\mu}-2L_{\chi_{1\mu}'}+i\pi
-\frac{2s}{s-\chi_1'}\left[ L_{s\mu}-L_{\chi_{1\mu}'} \right]
+sI_{134}\right],\nonumber\\
b_{134}&=&a_{134}-I_{134}=\frac{1}{s-\chi_1'}\left[-L_{s\mu}+i\pi
-\frac{2\chi_1'}{s-\chi_1'}\left[ L_{s\mu}-L_{\chi_{1\mu}'} \right]
+\chi_1'I_{134}\right],\nonumber\\
c_{134}&=&d_{134}=\frac{1}{s-\chi_1'}\left[L_{\chi_{1\mu}'}-i\pi
+\frac{2s}{s-\chi_1'}\left[ L_{s\mu}-L_{\chi_{1\mu}'} \right]
-sI_{134}\right].\\
&&\nonumber\\
a_{124}&=&I_{124}=\frac{1}{\chi_1'}\left[\frac{1}{2}L_{\chi_{1m}'}^2
-\zeta (2)-i\pi L_{\chi_{1m}'}\right],\
d_{124}=\frac{1}{\chi_{1m}'} \left[-L_{\chi_{1m}'} +i\pi \right],\nonumber\\
c_{124}&=&\frac{1}{\chi_1'}\left[
-\frac{1}{2}L_{\chi_{1m}'}^2 + L_{\chi_{1m}'} + \zeta (2)-2
+i\pi \left( L_{\chi_{1m}'}-1 \right)\right],\ b_{124}=0.\\
&&\nonumber\\
a_{123}&=&b_{123}-I_{123}=\frac{1}{s-s_1}\biggl[
\frac{1}{2}L_{s_m}^2 -\frac{1}{2}L_{s_{1m}}^2-L_{s_m}+L_{s_{1m}} \nonumber\\
&+&i\pi\left( L_{s_{1m}}-L_{s_m} \right)\biggr],\
b_{123}=-\frac{1}{s-s_1}\left[ L_{s_m}-L_{s_{1m}}\right],\
d_{123}=0,\nonumber\\
c_{123}&=&\frac{1}{s-s_1}\biggl[L_{s_{1m}}-2-i\pi+\frac{s}{s-s_1}\biggl[
-\frac{1}{2}L_{s_m}^2+\frac{1}{2}L_{s_{1m}}^2 \nonumber\\
&+&2L_{s_m}-2L_{s_{1m}}+i\pi\left( L_{s_m}-L_{s_{1m}} \right)\biggr] \biggr].
\end{eqnarray}

\subsection{Four-propagator integrals}
\subsubsection{Scalar}
\begin{eqnarray}
I_{1245}&=&\int_{0}^{1}\frac{\dd x\dd y}{\left[ xy\chi_1-\bar yt\right]
\left[ ym^2-\bar x\bar y \chi_1'\right]}=\frac{1}{\chi_1\chi_1'}\biggl[
-L_{\chi_{1m}}^2-L_{\chi_{1m}'}^2-L_{t_m}^2 \nonumber\\
&-&2L_{\chi_{1m}}L_{\chi_{1m}'}+2L_{\chi_{1m}}L_{t_m}
+2L_{\chi_{1m}'}L_t+4\zeta (2)+2i\pi\left( L_{\chi_{1m}}
+L_{\chi_{1m}'}-L_{t_m} \right)\biggr],\nonumber\\
I_{2345}&=&\frac{1}{t}\int_{0}^{1}\frac{\dd x}{{\cal P}_1^2}\left[
-\frac{1}{2}\ln\left(\frac{{\cal P}_1^2}{\lambda^2}\right)
+\ln\left( \frac{x\chi_1}{-t} \right)\right] \nonumber\\
&=&\frac{1}{s_1t}\biggl[
L_{s_{1\mu}}^2-L_{s_{1\mu}}L_{\lambda_\mu}-2L_{s_{1\mu}}L_{\chi_{1\mu}}
+2L_{s_{1\mu}}L_{t_\mu}-5\zeta (2) \nonumber\\
&+&i\pi\left( 2L_{\chi_{1\mu}}-2L_{t_\mu}-2L_{s_{1\mu}}
+L_{\lambda_\mu} \right)+\ln\frac{\mu}{m}\left(-2L_{\chi_{1\mu}}+2L_{t_\mu}
-L_{\lambda_\mu}\right) - \ln^2\frac{\mu}{m} \biggr],\nonumber\\
I_{1345}&=&\frac{1}{t}\int_{0}^{1}\frac{\dd x}{{\cal P}^2}\left[
-\frac{1}{2}\ln\left(\frac{{\cal P}^2}{\lambda^2}\right)
+\ln\left( \frac{-x\chi_1'}{-t} \right)\right]\nonumber\\
&=&\frac{1}{st}\biggl[
\frac{1}{2}L_{s_m}^2+\frac{1}{2}L_{s_\mu}^2 - L_{s_m}L_{\lambda_m}
-2L_{s_m}L_{\chi_{1m}'}+2L_{s_m}L_{t_m}+7\zeta (2) \nonumber\\
&+&i\pi\left( 2L_{\chi_{1m}'}-2L_{t_m}+L_{\lambda_m} \right)
-\ln^2\frac{\mu}{m}-\ln\frac{\mu}{m}\left(L_{\lambda_m}
+2L_{\chi_{1m}'}-2L_{t_m}-4i\pi \right)\biggr], \nonumber\\
I_{1235}&=&\frac{1}{\chi_1}\int_{0}^{1}\frac{\dd x}{{\cal P}^2}\left[
\frac{1}{2}\ln\left(\frac{{\cal P}^2}{\lambda^2}\right)
-\ln\left( \frac{{\cal P}_1^2}{x\chi_1} \right)\right]
=\frac{1}{s\chi_1}\Biggl[\frac{1}{2}L_{s_\mu}^2-\frac{1}{2}L_{s_m}^2
+\frac{1}{2}L_{s_{1m}}^2 \nonumber\\
&+&\frac{1}{2}L_{s_{1\mu}}^2
-2L_{s_m}L_{\chi_{1m}}+L_{s_m}L_{\lambda_m}-5\zeta (2)
+2\mbox{Li}_2\left( 1-\frac{s_1}{s}\right)+i\pi( -2L_{s_{1\mu}}\nonumber\\
&+&2L_{\chi_{1m}}-L_{\lambda_m})
+\ln^2\frac{\mu}{m}+\ln\frac{\mu}{m}\left(2L_{s_\mu}+2L_{\chi_{1m}}
-L_{\lambda_m} - 2i\pi \right) \Biggr],\nonumber\\
I_{1234}&=&\frac{1}{\chi_1'}\int_{0}^{1}\frac{\dd x}{{\cal P}_1^2}\left[
-\frac{1}{2}\ln\left(\frac{{\cal P}_1^2}{\lambda^2}\right)
+\ln\left( \frac{{\cal P}^2}{-x\chi_1'} \right)\right]
=\frac{1}{s_1\chi_1'}\Biggl[-\frac{1}{2}L_{s_\mu}^2-\frac{1}{2}L_{s_m}^2
+\frac{1}{2}L_{s_{1m}}^2 \nonumber\\
&-&\frac{1}{2}L_{s_{1\mu}}^2
+2L_{s_{1m}}L_{\chi_{1m}'}-L_{s_{1m}}L_{\lambda_m}-7\zeta (2)
-2\mbox{Li}_2\left( 1-\frac{s}{s_1}\right)+i\pi( 2L_{s_\mu}-2L_{s_{1m}}
\nonumber\\
&-&2L_{\chi_{1m}'}+L_{\lambda_m})
-\ln^2\frac{\mu}{m}+\ln\frac{\mu}{m}\left(-2L_{s_{1\mu}}-2L_{\chi_{1m}'}
+L_{\lambda_m}+4i\pi \right) \Biggr].
\end{eqnarray}

Useful integrals

\begin{eqnarray}
\int_{0}^{1}\frac{\dd x}{{\cal P}^2}
&=&\frac{2}{s}\left[-L_{s_m}+\ln\frac{\mu}{m}+i\pi\right],\nonumber\\
\int_{0}^{1}\frac{\dd x}{{\cal P}^2}\ln x
&=&\frac{1}{s}\left[\frac{1}{2}L_{s_\mu}^2-\zeta (2)-i\pi L_{s_\mu}\right],
\nonumber\\
\int_{0}^{1}\frac{\dd x}{{\cal P}^2}\ln \left( \frac{{\cal P}^2}{m^2}\right)
&=&\frac{1}{s}\left[-L_{s_m}^2+8\zeta (2)+2i\pi L_{s_m}+2\ln^2\frac{\mu}{m}
\right], \nonumber\\
\int_{0}^{1}\frac{\dd x}{{\cal P}^2}\ln \left( \frac{{\cal P}_1^2}{m^2}\right)
&=&\frac{1}{s}\Biggl[-\frac{1}{2}L_{s_{1m}}^2-\frac{1}{2}L_{s_{1\mu}}^2+8\zeta (2)
-2\mbox{Li}_2\left(1-\frac{s_1}{s} \right) \nonumber\\
&-&2\ln\frac{\mu}{m}L_{s_\mu}
+i\pi\left(L_{s_{1m}}+L_{s_{1\mu}}+2\ln\frac{\mu}{m}\right)\Biggr].
\end{eqnarray}

The vector, tensor four--propagator and pentagon integrals are the same as
in the case of equal fermion masses.

\section{Self--energy and real--photon vertex corrections }

In this section the masses are taken into account.
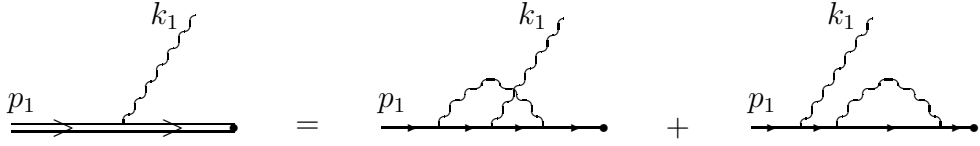
\begin{figure}[h]
\unitlength=2.10pt
\special{em:linewidth 0.4pt}
\linethickness{0.4pt}
\begin{picture}(194.00,34.33)
\put(20.00,13.00){\line(1,0){40.00}}
\put(60.00,14.33){\line(-1,0){40.00}}
\put(60.00,13.67){\circle*{1.33}}
\put(41.13,14.33){\oval(2.00,3.33)[lt]}
\put(41.13,17.67){\oval(2.00,3.33)[rb]}
\put(43.13,17.33){\oval(2.00,3.33)[lt]}
\put(43.13,20.67){\oval(2.00,3.33)[rb]}
\put(45.13,20.33){\oval(2.00,3.33)[lt]}
\put(45.13,23.67){\oval(2.00,3.33)[rb]}
\put(47.13,23.33){\oval(2.00,3.33)[lt]}
\put(47.13,26.67){\oval(2.00,3.33)[rb]}
\put(49.13,26.33){\oval(2.00,3.33)[lt]}
\put(49.13,29.67){\oval(2.00,3.33)[rb]}
\put(51.13,29.33){\oval(2.00,3.33)[lt]}
\put(51.13,32.67){\oval(2.00,3.33)[rb]}
\put(53.13,32.33){\oval(2.00,3.33)[lt]}
\put(73.33,13.67){\makebox(0,0)[cc]{$=$}}
\put(86.67,13.67){\line(1,0){40.00}}
\put(49.00,13.67){\makebox(0,0)[cc]{$>$}}
\put(29.33,13.67){\makebox(0,0)[cc]{$>$}}
\put(98.00,13.66){\oval(2.00,4.00)[lt]}
\put(114.66,13.66){\oval(2.00,4.00)[rt]}
\put(98.00,17.33){\oval(2.00,3.33)[rb]}
\put(114.66,17.33){\oval(2.00,3.33)[lb]}
\put(100.00,17.33){\oval(2.00,2.67)[lt]}
\put(112.66,17.33){\oval(2.00,2.67)[rt]}
\put(100.00,19.66){\oval(3.33,2.00)[rb]}
\put(112.66,19.66){\oval(3.33,2.00)[lb]}
\put(103.66,19.99){\oval(4.00,1.33)[lt]}
\put(108.99,19.99){\oval(4.00,1.33)[rt]}
\put(103.66,21.66){\oval(2.00,2.00)[rb]}
\put(108.99,21.66){\oval(2.00,2.00)[lb]}
\put(105.66,21.32){\oval(2.00,2.00)[lt]}
\put(106.99,21.32){\oval(2.00,2.00)[rt]}
\put(105.66,22.33){\line(1,0){1.33}}
\put(107.46,13.66){\oval(2.00,3.33)[lt]}
\put(107.46,17.00){\oval(2.00,3.33)[rb]}
\put(109.46,16.66){\oval(2.00,3.33)[lt]}
\put(109.46,20.00){\oval(2.00,3.33)[rb]}
\put(111.46,19.66){\oval(2.00,3.33)[lt]}
\put(111.46,23.00){\oval(2.00,3.33)[rb]}
\put(113.46,22.66){\oval(2.00,3.33)[lt]}
\put(113.46,26.00){\oval(2.00,3.33)[rb]}
\put(115.46,25.66){\oval(2.00,3.33)[lt]}
\put(115.46,29.00){\oval(2.00,3.33)[rb]}
\put(117.46,28.66){\oval(2.00,3.33)[lt]}
\put(117.46,32.00){\oval(2.00,3.33)[rb]}
\put(119.46,31.66){\oval(2.00,3.33)[lt]}
\put(121.67,13.67){\vector(1,0){1.00}}
\put(111.67,13.67){\vector(1,0){1.00}}
\put(102.67,13.67){\vector(1,0){1.00}}
\put(92.33,13.67){\vector(1,0){1.00}}
\put(126.67,13.67){\circle*{1.33}}
\put(140.00,13.67){\makebox(0,0)[cc]{$+$}}
\put(153.34,13.67){\line(1,0){40.00}}
\put(169.67,13.66){\oval(2.00,4.00)[lt]}
\put(186.33,13.66){\oval(2.00,4.00)[rt]}
\put(169.67,17.33){\oval(2.00,3.33)[rb]}
\put(186.33,17.33){\oval(2.00,3.33)[lb]}
\put(171.67,17.33){\oval(2.00,2.67)[lt]}
\put(184.33,17.33){\oval(2.00,2.67)[rt]}
\put(171.67,19.66){\oval(3.33,2.00)[rb]}
\put(184.33,19.66){\oval(3.33,2.00)[lb]}
\put(175.33,19.99){\oval(4.00,1.33)[lt]}
\put(180.66,19.99){\oval(4.00,1.33)[rt]}
\put(175.33,21.66){\oval(2.00,2.00)[rb]}
\put(180.66,21.66){\oval(2.00,2.00)[lb]}
\put(177.33,21.32){\oval(2.00,2.00)[lt]}
\put(178.66,21.32){\oval(2.00,2.00)[rt]}
\put(177.33,22.33){\line(1,0){1.33}}
\put(163.13,13.66){\oval(2.00,3.33)[lt]}
\put(163.13,17.00){\oval(2.00,3.33)[rb]}
\put(165.13,16.66){\oval(2.00,3.33)[lt]}
\put(165.13,20.00){\oval(2.00,3.33)[rb]}
\put(167.13,19.66){\oval(2.00,3.33)[lt]}
\put(167.13,23.00){\oval(2.00,3.33)[rb]}
\put(169.13,22.66){\oval(2.00,3.33)[lt]}
\put(169.13,26.00){\oval(2.00,3.33)[rb]}
\put(171.13,25.66){\oval(2.00,3.33)[lt]}
\put(171.13,29.00){\oval(2.00,3.33)[rb]}
\put(173.13,28.66){\oval(2.00,3.33)[lt]}
\put(173.13,32.00){\oval(2.00,3.33)[rb]}
\put(175.13,31.66){\oval(2.00,3.33)[lt]}
\put(157.00,13.67){\vector(1,0){1.00}}
\put(193.34,13.67){\circle*{1.33}}
\put(165.33,13.67){\vector(1,0){1.00}}
\put(178.67,13.67){\vector(1,0){1.00}}
\put(190.33,13.67){\vector(1,0){1.00}}
\put(21.90,18.10){\makebox(0,0)[cc]{$p_1$}}
\put(88.57,18.10){\makebox(0,0)[cc]{$p_1$}}
\put(155.24,18.10){\makebox(0,0)[cc]{$p_1$}}
\put(47.62,33.81){\makebox(0,0)[cc]{$k_1$}}
\put(113.81,33.81){\makebox(0,0)[cc]{$k_1$}}
\put(169.52,33.81){\makebox(0,0)[cc]{$k_1$}}
\end{picture}
\caption{\bf Self--energy and real--photon vertex corrections
to the incoming fermion}
\end{figure}

\begin{equation}
{\cal L}_1
=\left[
\frac{\not\!p_1-\not\!k_1+m}{-\chi_1}\Gamma_\mu e^\mu
+\Sigma (\not\!p_1-\not\!k_1)\frac{1}{(\not\!p_1-\not\!k_1-m)^2}\not\!e
\right]u(p_1)
\end{equation}
Using well known expressions for the off-shell vertex function $\Gamma$
and mass operator $\Sigma$, we obtain

\begin{equation}
{\cal L}_1
=\frac{\alpha}{2\pi}\left[
{\cal A}_1\left( \not\!e -\not\!k_1\frac{ep_1}{k_1p_1} \right)
+{\cal A}_2 \not\!k_1\!\!\not\!e
\right]u(p_1),
\end{equation}
where
\begin{eqnarray}
{\cal A}_1&=&-\frac{m}{2(\chi_1-m^2)}
\left[ 1-\frac{\chi_1}{\chi_1-m^2}L_{\chi_1} \right],\nonumber\\
{\cal A}_2&=&-\frac{1}{2(\chi_1-m^2)}
+\frac{2\chi_1^2-3m^2\chi_1+2m^4}{2\chi_1(\chi_1-m^2)^2}L_{\chi_1}
+\frac{m^2}{\chi_1^2}\biggl[-\Li\left( 1-\frac{\chi_1}{m^2} \right) \nonumber\\
&+&\zeta(2)\biggr].
\end{eqnarray}

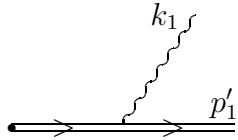
\begin{figure}[h]
\unitlength=2.10pt
\special{em:linewidth 0.4pt}
\linethickness{0.4pt}
\begin{picture}(126.67,34.33)
\put(86.67,13.00){\line(1,0){40.00}}
\put(126.67,14.33){\line(-1,0){40.00}}
\put(86.67,13.67){\circle*{1.33}}
\put(107.80,14.33){\oval(2.00,3.33)[lt]}
\put(107.80,17.67){\oval(2.00,3.33)[rb]}
\put(109.80,17.33){\oval(2.00,3.33)[lt]}
\put(109.80,20.67){\oval(2.00,3.33)[rb]}
\put(111.80,20.33){\oval(2.00,3.33)[lt]}
\put(111.80,23.67){\oval(2.00,3.33)[rb]}
\put(113.80,23.33){\oval(2.00,3.33)[lt]}
\put(113.80,26.67){\oval(2.00,3.33)[rb]}
\put(115.80,26.33){\oval(2.00,3.33)[lt]}
\put(115.80,29.67){\oval(2.00,3.33)[rb]}
\put(117.80,29.33){\oval(2.00,3.33)[lt]}
\put(117.80,32.67){\oval(2.00,3.33)[rb]}
\put(119.80,32.33){\oval(2.00,3.33)[lt]}
\put(115.67,13.67){\makebox(0,0)[cc]{$>$}}
\put(96.00,13.67){\makebox(0,0)[cc]{$>$}}
\put(114.29,33.81){\makebox(0,0)[cc]{$k_1$}}
\put(125.00,18.00){\makebox(0,0)[cc]{$p_1'$}}
\end{picture}
\caption{\bf Self--energy and real--photon vertex corrections
to the outgoing fermion }
\end{figure}

\begin{equation}
{\cal L}_1'
=\frac{\alpha}{2\pi}\bar u(p_1')\left[
{\cal B}_1\left( \not\!e -\not\!k_1\frac{ep_1'}{k_1p_1'} \right)
+{\cal B}_2 \not\!k_1\!\!\not\!e
\right],
\end{equation}
where
\begin{eqnarray}
{\cal B}_1&=&\frac{m}{2(\chi_1'+m^2)}
\left[ 1-\frac{\chi_1'}{\chi_1'+m^2}
\left( L_{\chi_1'}-i\pi \right)\right],\nonumber\\
{\cal B}_2&=&\frac{1}{2(\chi_1'+m^2)}
-\frac{2\chi_1'^2+3m^2\chi_1'+2m^4}{2\chi_1'(\chi_1'+m^2)^2}
\left( L_{\chi_1'}-i\pi \right)
+\frac{m^2}{\chi_1'^2}\biggl[ -\Li\left( 1+\frac{\chi_1'}{m^2} \right)\nonumber\\
&+&\zeta(2)\biggr].
\end{eqnarray}

\begin{figure}[h]
\unitlength=2.10pt
\special{em:linewidth 0.4pt}
\linethickness{0.4pt}
\begin{picture}(127.33,34.33)
\put(86.67,13.00){\line(1,0){40.00}}
\put(126.67,14.33){\line(-1,0){40.00}}
\put(126.67,13.67){\circle*{1.33}}
\put(107.80,14.33){\oval(2.00,3.33)[lt]}
\put(107.80,17.67){\oval(2.00,3.33)[rb]}
\put(109.80,17.33){\oval(2.00,3.33)[lt]}
\put(109.80,20.67){\oval(2.00,3.33)[rb]}
\put(111.80,20.33){\oval(2.00,3.33)[lt]}
\put(111.80,23.67){\oval(2.00,3.33)[rb]}
\put(113.80,23.33){\oval(2.00,3.33)[lt]}
\put(113.80,26.67){\oval(2.00,3.33)[rb]}
\put(115.80,26.33){\oval(2.00,3.33)[lt]}
\put(115.80,29.67){\oval(2.00,3.33)[rb]}
\put(117.80,29.33){\oval(2.00,3.33)[lt]}
\put(117.80,32.67){\oval(2.00,3.33)[rb]}
\put(119.80,32.33){\oval(2.00,3.33)[lt]}
\put(115.67,13.67){\makebox(0,0)[cc]{$<$}}
\put(96.00,13.67){\makebox(0,0)[cc]{$<$}}
\put(88.57,18.10){\makebox(0,0)[cc]{$-p_2$}}
\put(114.29,33.81){\makebox(0,0)[cc]{$k_1$}}
\end{picture}
\caption{\bf Self--energy and real--photon vertex corrections to the
incoming anti-fermion }
\end{figure}
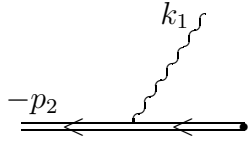

\begin{equation}
{\cal L}_2
=\frac{\alpha}{2\pi}\bar u(-p_2)\left[
{\cal C}_1\left( \not\!e -\not\!k_1\frac{ep_2}{k_1p_2} \right)
+{\cal C}_2 \not\!k_1\!\!\not\!e
\right],
\end{equation}
where
\begin{eqnarray}
{\cal C}_1&=&-\frac{m}{2(\chi_2-m^2)}
\left[ 1-\frac{\chi_2}{\chi_2-m^2}L_{\chi_2} \right],\nonumber\\
{\cal C}_2&=&-\frac{1}{2(\chi_2-m^2)}
+\frac{2\chi_2^2-3m^2\chi_2+2m^4}{2\chi_2(\chi_2-m^2)^2}L_{\chi_2}
+\frac{m^2}{\chi_2^2}\biggl[ -\Li\left( 1-\frac{\chi_2}{m^2} \right)\nonumber\\
&+&\zeta(2)\biggr].
\end{eqnarray}

\begin{figure}[h]
\unitlength=2.10pt
\special{em:linewidth 0.4pt}
\linethickness{0.4pt}
\begin{picture}(126.67,34.33)
\put(86.67,13.00){\line(1,0){40.00}}
\put(126.67,14.33){\line(-1,0){40.00}}
\put(86.67,13.67){\circle*{1.33}}
\put(107.80,14.33){\oval(2.00,3.33)[lt]}
\put(107.80,17.67){\oval(2.00,3.33)[rb]}
\put(109.80,17.33){\oval(2.00,3.33)[lt]}
\put(109.80,20.67){\oval(2.00,3.33)[rb]}
\put(111.80,20.33){\oval(2.00,3.33)[lt]}
\put(111.80,23.67){\oval(2.00,3.33)[rb]}
\put(113.80,23.33){\oval(2.00,3.33)[lt]}
\put(113.80,26.67){\oval(2.00,3.33)[rb]}
\put(115.80,26.33){\oval(2.00,3.33)[lt]}
\put(115.80,29.67){\oval(2.00,3.33)[rb]}
\put(117.80,29.33){\oval(2.00,3.33)[lt]}
\put(117.80,32.67){\oval(2.00,3.33)[rb]}
\put(119.80,32.33){\oval(2.00,3.33)[lt]}
\put(115.67,13.67){\makebox(0,0)[cc]{$<$}}
\put(96.00,13.67){\makebox(0,0)[cc]{$<$}}
\put(114.29,33.81){\makebox(0,0)[cc]{$k_1$}}
\put(125.00,18.00){\makebox(0,0)[cc]{$-p_2'$}}
\end{picture}
\caption{\bf Self--energy and real--photon vertex corrections to the outgoing
anti-fermion }
\end{figure}
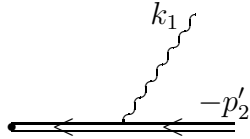

\begin{equation}
{\cal L}_2'
=\frac{\alpha}{2\pi}\left[
{\cal D}_1\left( \not\!e -\not\!k_1\frac{ep_2'}{k_1p_2'} \right)
+{\cal D}_2 \not\!k_1\!\!\not\!e
\right]u(-p_2'),
\end{equation}
where
\begin{eqnarray}
{\cal D}_1&=&\frac{m}{2(\chi_2'+m^2)}
\left[ 1-\frac{\chi_2'}{\chi_2'+m^2}
\left( L_{\chi_2'}-i\pi \right)\right],\nonumber\\
{\cal D}_2&=&\frac{1}{2(\chi_2'+m^2)}
-\frac{2\chi_2'^2+3m^2\chi_2'+2m^4}{2\chi_2'(\chi_2'+m^2)^2}
\left( L_{\chi_2'}-i\pi \right) \nonumber\\
&+&\frac{m^2}{\chi_2'^2}\biggl[ -\Li\left( 1+\frac{\chi_2'}{m^2} \right)
+\zeta(2)\biggr].
\end{eqnarray}

\section{Heavy--photon vertex diagrams }

\subsection{Notations}

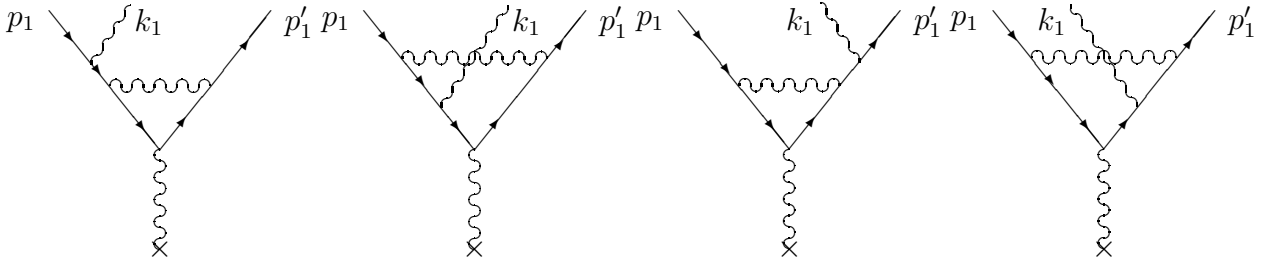
\begin{figure}[h]
\unitlength=2.10pt
\special{em:linewidth 0.4pt}
\linethickness{0.4pt}
\begin{picture}(227.15,72.76)
\put(32.14,28.00){\oval(2.00,2.00)[l]}
\put(32.14,30.00){\oval(2.00,2.00)[r]}
\put(32.14,32.00){\oval(2.00,2.00)[l]}
\put(32.14,34.00){\oval(2.00,2.00)[r]}
\put(32.14,36.00){\oval(2.00,2.00)[l]}
\put(32.14,38.00){\oval(2.00,2.00)[r]}
\put(32.14,40.00){\oval(2.00,2.00)[l]}
\put(32.14,42.00){\oval(2.00,2.00)[r]}
\put(32.14,44.00){\oval(2.00,2.00)[l]}
\put(12.14,70.00){\line(4,-5){20.00}}
\put(32.14,45.00){\line(4,5){20.00}}
\put(24.14,56.51){\oval(2.00,2.33)[t]}
\put(26.14,56.67){\oval(2.00,2.67)[b]}
\put(28.14,56.51){\oval(2.00,2.33)[t]}
\put(30.14,56.67){\oval(2.00,2.67)[b]}
\put(32.14,56.51){\oval(2.00,2.33)[t]}
\put(34.14,56.67){\oval(2.00,2.67)[b]}
\put(36.14,56.51){\oval(2.00,2.33)[t]}
\put(38.14,56.67){\oval(2.00,2.67)[b]}
\put(40.14,56.51){\oval(2.00,2.33)[t]}
\put(32.14,27.00){\makebox(0,0)[cc]{$\times$}}
\put(14.80,66.67){\vector(3,-4){1.67}}
\put(27.80,50.67){\vector(3,-4){1.67}}
\put(88.81,28.00){\oval(2.00,2.00)[l]}
\put(88.81,30.00){\oval(2.00,2.00)[r]}
\put(88.81,32.00){\oval(2.00,2.00)[l]}
\put(88.81,34.00){\oval(2.00,2.00)[r]}
\put(88.81,36.00){\oval(2.00,2.00)[l]}
\put(88.81,38.00){\oval(2.00,2.00)[r]}
\put(88.81,40.00){\oval(2.00,2.00)[l]}
\put(88.81,42.00){\oval(2.00,2.00)[r]}
\put(88.81,44.00){\oval(2.00,2.00)[l]}
\put(68.81,70.00){\line(4,-5){20.00}}
\put(88.81,45.00){\line(4,5){20.00}}
\put(76.81,61.51){\oval(2.00,2.33)[t]}
\put(78.81,61.67){\oval(2.00,2.67)[b]}
\put(80.81,61.51){\oval(2.00,2.33)[t]}
\put(82.81,61.67){\oval(2.00,2.67)[b]}
\put(84.81,61.51){\oval(2.00,2.33)[t]}
\put(86.81,61.67){\oval(2.00,2.67)[b]}
\put(88.81,61.51){\oval(2.00,2.33)[t]}
\put(90.81,61.67){\oval(2.00,2.67)[b]}
\put(92.81,61.51){\oval(2.00,2.33)[t]}
\put(94.81,61.67){\oval(2.00,2.67)[b]}
\put(96.81,61.51){\oval(2.00,2.33)[t]}
\put(98.81,61.67){\oval(2.00,2.67)[b]}
\put(100.81,61.51){\oval(2.00,2.33)[t]}
\put(83.81,52.66){\oval(2.00,3.33)[lt]}
\put(83.81,56.00){\oval(2.00,3.33)[rb]}
\put(85.81,55.66){\oval(2.00,3.33)[lt]}
\put(85.81,59.00){\oval(2.00,3.33)[rb]}
\put(87.81,58.66){\oval(2.00,3.33)[lt]}
\put(87.81,62.00){\oval(2.00,3.33)[rb]}
\put(89.81,61.66){\oval(2.00,3.33)[lt]}
\put(89.81,65.00){\oval(2.00,3.33)[rb]}
\put(91.81,64.66){\oval(2.00,3.33)[lt]}
\put(91.81,68.00){\oval(2.00,3.33)[rb]}
\put(93.81,67.66){\oval(2.00,3.33)[lt]}
\put(93.81,71.00){\oval(2.00,3.33)[rb]}
\put(88.81,27.00){\makebox(0,0)[cc]{$\times$}}
\put(71.47,66.67){\vector(3,-4){1.67}}
\put(78.14,58.67){\vector(3,-4){1.67}}
\put(84.47,50.67){\vector(3,-4){1.67}}
\put(20.80,60.33){\oval(2.00,3.33)[lt]}
\put(20.80,63.67){\oval(2.00,3.33)[rb]}
\put(22.80,63.33){\oval(2.00,3.33)[lt]}
\put(22.80,66.67){\oval(2.00,3.33)[rb]}
\put(24.80,66.33){\oval(2.00,3.33)[lt]}
\put(24.80,69.67){\oval(2.00,3.33)[rb]}
\put(26.80,69.33){\oval(2.00,3.33)[lt]}
\put(19.80,60.67){\vector(3,-4){1.67}}
\put(46.47,62.67){\vector(3,4){1.67}}
\put(35.14,48.67){\vector(3,4){1.67}}
\put(103.14,62.67){\vector(3,4){1.67}}
\put(91.47,48.34){\vector(3,4){1.67}}
\put(7.14,68.00){\makebox(0,0)[cc]{$p_1$}}
\put(63.81,68.00){\makebox(0,0)[cc]{$p_1$}}
\put(57.14,68.00){\makebox(0,0)[cc]{$p_1'$}}
\put(113.81,68.00){\makebox(0,0)[cc]{$p_1'$}}
\put(30.14,68.00){\makebox(0,0)[cc]{$k_1$}}
\put(98.14,68.00){\makebox(0,0)[cc]{$k_1$}}
\put(145.48,28.10){\oval(2.00,2.00)[l]}
\put(145.48,30.10){\oval(2.00,2.00)[r]}
\put(145.48,32.10){\oval(2.00,2.00)[l]}
\put(145.48,34.10){\oval(2.00,2.00)[r]}
\put(145.48,36.10){\oval(2.00,2.00)[l]}
\put(145.48,38.10){\oval(2.00,2.00)[r]}
\put(145.48,40.10){\oval(2.00,2.00)[l]}
\put(145.48,42.10){\oval(2.00,2.00)[r]}
\put(145.48,44.10){\oval(2.00,2.00)[l]}
\put(125.48,70.10){\line(4,-5){20.00}}
\put(145.48,45.10){\line(4,5){20.00}}
\put(137.48,56.61){\oval(2.00,2.33)[t]}
\put(139.48,56.77){\oval(2.00,2.67)[b]}
\put(141.48,56.61){\oval(2.00,2.33)[t]}
\put(143.48,56.77){\oval(2.00,2.67)[b]}
\put(145.48,56.61){\oval(2.00,2.33)[t]}
\put(147.48,56.77){\oval(2.00,2.67)[b]}
\put(149.48,56.61){\oval(2.00,2.33)[t]}
\put(151.48,56.77){\oval(2.00,2.67)[b]}
\put(153.48,56.61){\oval(2.00,2.33)[t]}
\put(145.48,27.10){\makebox(0,0)[cc]{$\times$}}
\put(128.14,66.77){\vector(3,-4){1.67}}
\put(141.14,50.77){\vector(3,-4){1.67}}
\put(202.15,28.10){\oval(2.00,2.00)[l]}
\put(202.15,30.10){\oval(2.00,2.00)[r]}
\put(202.15,32.10){\oval(2.00,2.00)[l]}
\put(202.15,34.10){\oval(2.00,2.00)[r]}
\put(202.15,36.10){\oval(2.00,2.00)[l]}
\put(202.15,38.10){\oval(2.00,2.00)[r]}
\put(202.15,40.10){\oval(2.00,2.00)[l]}
\put(202.15,42.10){\oval(2.00,2.00)[r]}
\put(202.15,44.10){\oval(2.00,2.00)[l]}
\put(182.15,70.10){\line(4,-5){20.00}}
\put(202.15,45.10){\line(4,5){20.00}}
\put(190.15,61.61){\oval(2.00,2.33)[t]}
\put(192.15,61.77){\oval(2.00,2.67)[b]}
\put(194.15,61.61){\oval(2.00,2.33)[t]}
\put(196.15,61.77){\oval(2.00,2.67)[b]}
\put(198.15,61.61){\oval(2.00,2.33)[t]}
\put(200.15,61.77){\oval(2.00,2.67)[b]}
\put(202.15,61.61){\oval(2.00,2.33)[t]}
\put(204.15,61.77){\oval(2.00,2.67)[b]}
\put(206.15,61.61){\oval(2.00,2.33)[t]}
\put(208.15,61.77){\oval(2.00,2.67)[b]}
\put(210.15,61.61){\oval(2.00,2.33)[t]}
\put(212.15,61.77){\oval(2.00,2.67)[b]}
\put(214.15,61.61){\oval(2.00,2.33)[t]}
\put(202.15,27.10){\makebox(0,0)[cc]{$\times$}}
\put(184.81,66.77){\vector(3,-4){1.67}}
\put(191.48,58.77){\vector(3,-4){1.67}}
\put(197.81,50.77){\vector(3,-4){1.67}}
\put(133.14,60.77){\vector(3,-4){1.67}}
\put(159.81,62.77){\vector(3,4){1.67}}
\put(148.48,48.77){\vector(3,4){1.67}}
\put(216.48,62.77){\vector(3,4){1.67}}
\put(204.81,48.44){\vector(3,4){1.67}}
\put(120.48,68.10){\makebox(0,0)[cc]{$p_1$}}
\put(177.15,68.10){\makebox(0,0)[cc]{$p_1$}}
\put(170.48,68.10){\makebox(0,0)[cc]{$p_1'$}}
\put(227.15,68.10){\makebox(0,0)[cc]{$p_1'$}}
\put(147.48,68.10){\makebox(0,0)[cc]{$k_1$}}
\put(192.81,68.10){\makebox(0,0)[cc]{$k_1$}}
\put(197.15,67.77){\oval(2.00,3.33)[rt]}
\put(197.15,71.10){\oval(2.00,3.33)[lb]}
\put(199.15,64.77){\oval(2.00,3.33)[rt]}
\put(199.15,68.10){\oval(2.00,3.33)[lb]}
\put(201.15,61.77){\oval(2.00,3.33)[rt]}
\put(201.15,65.10){\oval(2.00,3.33)[lb]}
\put(203.15,58.77){\oval(2.00,3.33)[rt]}
\put(203.15,62.10){\oval(2.00,3.33)[lb]}
\put(205.15,55.77){\oval(2.00,3.33)[rt]}
\put(205.15,59.10){\oval(2.00,3.33)[lb]}
\put(207.15,52.77){\oval(2.00,3.33)[rt]}
\put(207.15,56.10){\oval(2.00,3.33)[lb]}
\put(153.15,66.77){\oval(2.00,3.33)[rt]}
\put(153.15,70.10){\oval(2.00,3.33)[lb]}
\put(155.15,63.77){\oval(2.00,3.33)[rt]}
\put(155.15,67.10){\oval(2.00,3.33)[lb]}
\put(157.15,60.77){\oval(2.00,3.33)[rt]}
\put(157.15,64.10){\oval(2.00,3.33)[lb]}
\put(151.15,69.77){\oval(2.00,3.33)[rt]}
\end{picture}
\caption{\bf Heavy--photon vertex diagrams with real--photon emission}
\end{figure}

\begin{equation}
J^{1,\mu,\mu\nu}_{ijkl}
\equiv\int\frac{\dd^4k}{i\pi^2}\frac{1,k^{\mu},k^{\mu}k^{\nu}}{(i)(j)(k)(l)}
\end{equation}

\begin{eqnarray}
(0)&=&k^2-\lambda^2,\ (1)=(p_1-k)^2-m^2,\ (2)=(p_1'-k)^2-m^2,\nonumber\\
(q)&=&(p_1-k_1-k)^2-m^2.
\end{eqnarray}

\subsection{Two-propagator integrals}
\subsubsection{Scalar}

\begin{eqnarray}
J_{01}&=&J_{02}=L_{\Lambda}+1,\
J_{12}=L_{\Lambda}-L_{t_1}+1,\
J_{0q}=L_{\Lambda}+1-L_{\chi_1},\nonumber\\
J_{1q}&=&L_{\Lambda}-1,\
J_{2q}=L_{\Lambda}-L_t+1.
\end{eqnarray}

\subsubsection{Vector}

\begin{eqnarray}
J^{\mu}_{01}&=&p_1^\mu
\left[ \frac{1}{2}L_{\Lambda} -\frac{1}{4}\right],\
J^{\mu}_{02}= p_1'^\mu
\left[ \frac{1}{2}L_{\Lambda} -\frac{1}{4}\right],\
J^{\mu}_{1q}= \left(p_1-\frac{1}{2}k_1\right)^\mu
\left[ L_{\Lambda} -\frac{3}{2} \right],\nonumber\\
J^{\mu}_{12}&=&(p_1 + p_1')^\mu
\left[ \frac{1}{2}L_{\Lambda}- \frac{1}{2}L_{t_1}
+\frac{1}{4}\right],\
J^{\mu}_{0q}= (p_1-k_1)^\mu
\left[ \frac{1}{2}L_{\Lambda}+\frac{1}{4}
- \frac{1}{2}L_{\chi_1} \right] ,\nonumber\\
J^{\mu}_{2q}&=&(p_1'+p_1-k_1)^\mu
\left[\frac{1}{2}L_{\Lambda} +\frac{1}{4}-\frac{1}{2}L_t\right].
\end{eqnarray}

\subsection{Three-propagator integrals}
\subsubsection{Scalar}

\begin{eqnarray}
J_{012}&=&\frac{1}{2t_1}
\left[ -2L_{\lambda}L_{t_1}+ L_{t_1}^2 - 2\zeta(2) \right] ,\
J_{12q}=\frac{1}{2(\chi_1'-\chi_1)}
\left[ L_t^2 - L_{t_1}^2 \right] ,\nonumber\\
J_{01q}&=& -\frac{1}{\chi_1}
\left[ -\Li\left( 1 - \frac{\chi_1}{m^2} \right) +\zeta(2) \right],\nonumber\\
J_{02q}&=&\frac{1}{\chi_1'+t_1}
\left[ L_t\left( L_t - L_{\chi_1} \right)
+ \frac{1}{2}\left( L_t - L_{\chi_1} \right)^2
+2\Li\left( 1 + \frac{\chi_1}{t}\right) \right].
\end{eqnarray}

\subsubsection{Vector}

Parameterization
\begin{equation}
J^{\mu}_{ijk}=a_{ijk}p_1^{\mu}+b_{ijk}p_1'^{\mu}
+c_{ijk}q^{\mu}.
\end{equation}

\begin{eqnarray}
a_{012}&=&b_{012}=\frac{1}{t_1}L_{t_1},\ c_{012}=0,\nonumber\\
&& \nonumber\\
a_{01q}&=&\frac{1}{\chi_1}
\left[ \chi_1J_{01q} + 2L_{\chi_1} - 2 \right],\
b_{01q}=-c_{01q}=\frac{1}{\chi_1}\left[ - L_{\chi_1} + 2 \right],\nonumber\\
&& \nonumber\\
a_{02q}&=&0,\ b_{02q}=\frac{\chi_1}{\chi_1'+t_1}J_{02q}
+2\frac{t}{(\chi_1'+t_1)^2}L_{t}
-\frac{t-\chi_1}{(\chi_1'+t_1)^2}L_{\chi_1},\nonumber\\
c_{02q}&=&-\frac{1}{\chi_1'+t_1}L_{t}+\frac{1}{\chi_1'+t_1}L_{\chi_1},\nonumber\\
&& \nonumber\\
a_{12q}&=&\frac{t}{\chi_1'-\chi_1}J_{12q}
+\frac{t+t_1}{(\chi_1'-\chi_1)^2}L_{t_1}
-2\frac{t}{(\chi_1'-\chi_1)^2}L_t+\frac{2}{\chi_1'-\chi_1},\nonumber\\
b_{12q}&=&J_{12q}-a_{12q},\nonumber\\
c_{12q}&=&\frac{t_1}{\chi_1'-\chi_1}J_{12q}
+2\frac{t_1}{(\chi_1'-\chi_1)^2}L_{t_1}-\frac{t+t_1}{(\chi_1'-\chi_1)^2}L_t
+\frac{2}{\chi_1'-\chi_1}.
\end{eqnarray}

\subsubsection{Tensor}

\begin{eqnarray}
J^{\mu\nu}_{ijk}&=&g^T_{ijk}g^{\mu\nu}+a^T_{ijk}p_1^{\mu}p_1^{\nu}
+b^T_{ijk}p_1'^{\mu}p_1'^{\nu}+c^T_{ijk}q^{\mu}q^{\nu}\nonumber\\
&+&\alpha^T_{ijk}\{ p_1^{\mu}p_1'^{\nu} \}
+\beta^T_{ijk}\{ p_1^{\mu}q^{\nu} \}
+\gamma^T_{ijk}\{ p_1'^{\mu}q^{\nu} \}.
\end{eqnarray}

\begin{eqnarray}
g^T_{012}
&=&\frac{1}{4}L_{\Lambda}-\frac{1}{4}L_{t_1}+\frac{3}{8},\
a^T_{012}=b^T_{012}=\frac{1}{2t_1}L_{t_1}-\frac{1}{2t_1},\nonumber\\
\alpha^T_{012}&=&\frac{1}{2t_1},\ c^T_{012}=\beta^T_{012}=\gamma^T_{012}=0.\\
&& \nonumber\\
g^T_{01q}&=&-\frac{1}{4}L_{\chi_1}+\frac{1}{4}L_{\Lambda}+\frac{3}{8},\
a^T_{01q}=J_{01q}+\frac{3}{\chi_1}L_{\chi_1}
-\frac{9}{2\chi_1},\nonumber\\
b^T_{01q}&=&c^T_{01q}=-\gamma^T_{01q}=-\frac{1}{2\chi_1}L_{\chi_1}
+\frac{1}{\chi_1},\
\beta^T_{01q}=-\alpha^T_{01q}
=\frac{1}{2\chi_1}L_{\chi_1}-\frac{3}{2\chi_1}.\\
&& \nonumber\\
g^T_{02q}&=&-\frac{1}{4}\frac{\chi_1}{\chi_1'+t_1}L_{\chi_1}
-\frac{1}{4}\frac{t}{(\chi_1'+t_1)}L_t +\frac{1}{4}L_{\Lambda}
+\frac{3}{8} ,\nonumber\\
b^T_{02q}&=&\left[-\frac{\chi_1(t-\chi_1)}{(\chi_1'+t_1)^3}
-\frac{1}{2}\frac{(t^2+2t\chi_1-\chi_1^2)}
{(\chi_1'+t_1)^3} \right] L_{\chi_1}+\frac{t(t+4\chi_1)}{(\chi_1'+t_1)^3}L_t
\nonumber\\
&+&\frac{t-\chi_1}{2(\chi_1'+t_1)^2}
+\frac{\chi_1^2}{(\chi_1'+t_1)^2}J_{02q},\nonumber\\
c^T_{02q}&=&-\frac{1}{2}\frac{1}{\chi_1'+t_1}L_{\chi_1}
+\frac{1}{2(\chi_1'+t_1)}L_t ,\
a^T_{02q}=\alpha^T_{02q}=\beta^T_{02q}=0,\nonumber\\
\gamma^T_{02q}&=&\frac{t+2\chi_1}{2(\chi_1'+t_1)^2}L_{\chi_1}
-\frac{t+2\chi_1}{2(\chi_1'+t_1)^2}L_t-\frac{1}{2(\chi_1'+t_1)}.\\
&& \nonumber\\
g^T_{12q}&=&\frac{1}{4}\frac{t_1}{\chi_1'-\chi_1}L_{t_1}
-\frac{1}{4}\frac{t}{\chi_1'-\chi_1}L_t
+\frac{1}{4}L_{\Lambda}+\frac{3}{8},\nonumber\\
a^T_{12q}&=&\frac{t^2}{(\chi_1'-\chi_1)^2}J_{12q}
+\frac{(3t^2+4t_1t-t_1^2)}{2(\chi_1'-\chi_1)^3}L_{t_1}
-\frac{3t^2}{(\chi_1'-\chi_1)^3}L_t
+\frac{4t-t_1}{(\chi_1'-\chi_1)^2},\nonumber\\
b^T_{12q}&=&\frac{t_1^2}{(\chi_1'-\chi_1)^2}J_{12q}
+\frac{(-t^2+4t_1t+3t_1^2)}{2(\chi_1'-\chi_1)^3}L_{t_1}
+\frac{t(t-4t_1)}{(\chi_1'-\chi_1)^3}L_t
+\frac{3t_1}{(\chi_1'-\chi_1)^2},\nonumber\\
c^T_{12q}&=&\frac{t_1^2}{(\chi_1'-\chi_1)^2}J_{12q}
+\frac{3t_1^2}{(\chi_1'-\chi_1)^3}L_{t_1}
+\frac{(t^2-4t_1t-3t_1^2)}{2(\chi_1'-\chi_1)^3}L_t
+\frac{(4t_1-t)}{(\chi_1'-\chi_1)^2},\nonumber\\
\alpha^T_{12q}&=&-\frac{t_1t}{(\chi_1'-\chi_1)^2}J_{12q}
-\frac{(t^2+4t_1t+t_1^2)}{2(\chi_1'-\chi_1)^3}L_{t_1}
+\frac{t(t+2t_1)}{(\chi_1'-\chi_1)^3}L_t
-\frac{(2t+t_1)}{(\chi_1'-\chi_1)^2},\nonumber\\
\beta^T_{12q}&=&\frac{t_1t}{(\chi_1'-\chi_1)^2}J_{12q}
+\frac{t_1(5t+t_1)}{2(\chi_1'-\chi_1)^3}L_{t_1}
-\frac{t(t+5t_1)}{2(\chi_1'-\chi_1)^3}L_t
+\frac{3}{2}\frac{t+t_1}{(\chi_1'-\chi_1)^2},\nonumber\\
\gamma^T_{12q}&=&-\frac{t_1^2}{(\chi_1'-\chi_1)^2}J_{12q}
-\frac{t_1(t+5t_1)}{2(\chi_1'-\chi_1)^3}L_{t_1}
+\frac{-t^2+5t_1t+2t_1^2}{2(\chi_1'-\chi_1)^3}L_t \nonumber\\
&+&\frac{t-7t_1}{2(\chi_1'-\chi_1)^2}.
\end{eqnarray}

\subsection{Four-propagator integrals}

\subsubsection{Scalar}

\begin{equation}
J_{012q}=-\frac{1}{\chi_1t_1}\left[ -L_{\lambda}L_{t_1}
+2L_{t_1}L_{\chi_1}-L_{t}^2 - 2\Li\left( 1-\frac{t}{t_1} \right)
-\zeta(2) \right].
\end{equation}

\subsubsection{Vector}

Parameterization
\begin{equation}
J^{\mu}_{012q}=a_{012q}p_1^\mu+b_{012q}p_1'^\mu+c_{012q}q^\mu.
\end{equation}

\begin{eqnarray}
a_{012q}
&=&\frac{1}{d}
\bigl[
-(t_1\chi_1+t\chi_1')J_{12q}
+(\chi_1'+t_1)^2J_{02q}
-\chi_1(\chi_1'-t_1)J_{01q}\nonumber\\
&-&t_1(\chi_1'+t_1)(J_{012}+\chi_1J_{012q})
\bigr],\nonumber\\
b_{012q}
&=&\frac{1}{d}
\bigl[
(t_1\chi_1'+t\chi_1)J_{12q} - (\chi_1\chi_1'+t_1t)J_{02q}
-\chi_1(t_1-\chi_1)J_{01q}\nonumber\\
&+&t_1(t_1-\chi_1)(J_{012}+\chi_1J_{012q})
\bigr],\nonumber\\
c_{012q}
&=&\frac{1}{d}
\bigl[
-t_1(\chi_1+\chi_1')J_{12q}
+t_1(\chi_1'+t_1)J_{02q}
+\chi_1t_1J_{01q}\nonumber\\
&-&t_1^2(J_{012}+\chi_1J_{012q})\bigr].
\end{eqnarray}
where $d=-2t_1\chi_1\chi_1'$.

\subsubsection{Tensor}

Parameterization
\begin{eqnarray}
J^{\mu\nu}_{012q}&=&g^T_{012q}g^{\mu\nu}+a^T_{012q}p_1^{\mu}p_1^{\nu}
+b^T_{012q}p_1'^{\mu}p_1'^{\nu}+c^T_{012q}q^{\mu}q^{\nu}\nonumber\\
&+&\alpha^T_{012q}\{ p_1^{\mu}p_1'^{\nu} \}
+\beta^T_{012q}\{ p_1^{\mu}q^{\nu} \}
+\gamma^T_{012q}\{ p_1'^{\mu}q^{\nu} \}.
\end{eqnarray}

\begin{eqnarray}
g^T_{012q}
&=&\frac{1}{2}\left[ J_{12q}-\chi_1c_{012q} \right],\nonumber\\
a^T_{012q}
&=&\frac{1}{d}
\bigl[ (\chi_1'+t_1)^2(J_{12q}-\chi_1c_{012q})
-(\chi_1t_1 + \chi_1't  )a_{12q}\nonumber\\
&-&\chi_1(\chi_1'-t_1)a_{01q}
-t_1(\chi_1'+t_1)(a_{012}+\chi_1a_{012q}) \bigr],\nonumber\\
b^T_{012q}
&=&\frac{1}{d} \bigl[(t_1-\chi_1)^2(J_{12q}-\chi_1c_{012q})
+(\chi_1't_1 + \chi_1t  )b_{12q}\nonumber\\
&-&(t_1t+\chi_1\chi_1')b_{02q}
+\chi_1(\chi_1-t_1)b_{01q}\nonumber\\
&+&t_1(t_1-\chi_1)(a_{012}+\chi_1b_{012q})
\bigr],\nonumber\\
c^T_{012q}
&=&\frac{1}{d} \bigl[ t_1^2(J_{12q}-2\chi_1c_{012q})
-t_1(\chi_1' + \chi_1  )c_{12q}\nonumber\\
&-&t_1\chi_1b_{01q}+t_1(t_1+\chi_1')c_{02q} \bigr],\nonumber\\
\alpha^T_{012q}
&=&\frac{1}{d} \bigl[
-( t_1t + \chi_1\chi_1' )(J_{12q}-\chi_1c_{012q})
+(\chi_1't_1 + \chi_1t  )a_{12q}\nonumber\\
&+&\chi_1(\chi_1-t_1)a_{01q}
+t_1(t_1-\chi_1)(a_{012}+\chi_1a_{012q}) \bigr],\nonumber\\
\beta^T_{012q}
&=&\frac{1}{d} \bigl[
t_1(t_1+\chi_1')(J_{12q}-2\chi_1c_{012q})
-(\chi_1t_1 + \chi_1't  )c_{12q}\nonumber\\
&+&(t_1+\chi_1')^2c_{02q}
+\chi_1(\chi_1'-t_1)b_{01q} \bigr],\nonumber\\
\gamma^T_{012q}
&=&\frac{1}{d} \bigl[
t_1(\chi_1-t_1)(J_{12q}-2\chi_1c_{012q})
+(\chi_1't_1 + \chi_1t  )c_{12q}\nonumber\\
&-&( t_1t+\chi_1'\chi_1 )c_{02q}
-\chi_1(\chi_1-t_1)b_{01q}
\bigr].
\end{eqnarray}

\end{document}